\documentclass{article}%
\usepackage{amsmath}
\usepackage{amsfonts}
\usepackage{amssymb}
\usepackage{graphicx}%
\begin{document}

\title{A Comment on Emergent Gravity}
\author{Waldyr A. Rodrigues Jr.\\Institute of Mathematics, Statistics and Scientific Computation\\IMECC-UNICAMP, CP 6065\\13083-970 Campinas, SP, Brazil\\walrod@ime.unicamp.br, }
\date{15 February 2006}
\maketitle
\tableofcontents

\begin{abstract}
This paper is a set of notes that we wrote concerning the first version of
\textit{Emergent Gravity} {\small [gr-qc/0602022]}. It is our version of an
exercise that we proposed to some of our students. The idea was to find
\textit{mathematical} errors and inconsistencies on some recent articles
published in scientific journals and in the arXiv, and we did.

\end{abstract}

\section{Introduction}

This paper is a set of notes that we wrote as a guide to a query that was
proposed to some of our students: find mathematical errors and inconsistencies
on some recent Physics papers published in scientific journals and/or posted
in the arXiv using jargon of higher Mathematics. The paper here analyzed is
the \textit{first version} of Emergent Gravity \cite{sarfatti}. Other papers
will be analyzed elsewhere (see, e.g., \cite{rodoliv2006}, where we criticise
\cite{car})

The reader is here informed that we sent the notes to the author of
\cite{sarfatti}, which used them to prepare new versions of his
paper\footnote{On February 27, 2006 we already have six versions.}. Originally
it was not our intention to post the notes but we changed our mind due to the
following reasons:

(i) our believe that notes may be eventually useful to many students and researchers;

(ii) because in the sixth `improved version' it is written in the comments of
the article: \ `6th draft due to math corrections by Prof Waldyr Rodrigues Jr
UNICAMP and new empirical information from UCLA Dark Matter 2006 Conference.

Well, unfortunately despite the fact that in the `improved versions' some of
the wrong mathematical statements of the first version have been deleted,
there are in our opinion new ones which need to be corrected\footnote{This
will be discussed elsewhere.}. We have no responsibility for any one of the
versions of that paper, we have not endorsed the paper for the arXiv.

\section{Some Necessary Preliminaries}

\subsection{Tetrads and Cotetrads}

\textbf{1.} The Collins Dictionary of Mathematics \cite{collins} defines the
word \textit{tetrad }as a set or sequence of four elements\textit{. }And
indeed, the prexif tetra comes from the Greek word for the number four. The
meaning of tetrads (and cotetrads) in differential geometry will be explained
next \cite{roqui} and that meaning must be kept in mind, specially in the
discussion in Section 7.

\textbf{2. }Let $M$ be a $4$-dimensional manifold equipped with a
\textit{Lorentzian} metric tensor field $\mathbf{g\in}\sec T_{0}^{2}M$. In the
differential geometry of $M$ the word tetrad is used to denominate a set of
\textit{four orthonormal} vector fields $\{\mathbf{e}_{\mathbf{a}}\}$ defined
in $U\subset M.$ We code this information writing $\mathbf{e}_{\mathbf{a}}%
\in\sec TU\subset\sec TM$, $\mathbf{a}=0,1,2,3$ and
\begin{equation}
\mathbf{g}(\mathbf{e}_{\mathbf{a}},\mathbf{e}_{\mathbf{b}})=\eta_{\mathbf{ab}%
}:=\mathrm{diag}(1,-1,-1,-1) \label{1}%
\end{equation}

\textbf{3.} Another way to code the above information is by writing that the
set $\{\mathbf{e}_{\mathbf{a}}\}$ $\in\sec\mathbf{P}_{\mathrm{SO}_{1,3}^{e}%
}U\subset\sec\mathbf{P}_{\mathrm{SO}_{1,3}^{e}}M$, i.e., set $\{\mathbf{e}%
_{\mathbf{a}}\}$ is a section of the orthonormal frame bundle, which is a
principal bundle with structural group \textrm{SO}$_{13}^{e}$. \ For details,
please consult, e.g., \cite{rodol}. Then the \textit{set} $\{\mathbf{e}%
_{\mathbf{a}}\}$ is also called an \textit{orthonormal (moving) frame}. It is
eventually important to recall a \textit{classical result} (see,
e.g.,\cite{choquet,moro,naka,rodol}) that for a $4$-dimensional Lorentzian
manifold to admit spinor fields $\mathbf{P}_{\mathrm{SO}_{1,3}^{e}}M$ must be
\textit{trivial}, i.e., must have global sections.

\textbf{4. }Sometimes it is useful to consider a set of \textit{vector fields}
$\{\mathbf{e}^{\mathbf{a}}\}\sec\mathbf{P}_{\mathrm{SO}_{1,3}^{e}}U\subset
\sec\mathbf{P}_{\mathrm{SO}_{1,3}^{e}}M$ such that
\begin{equation}
\mathbf{g}(\mathbf{e}^{\mathbf{a}},\mathbf{e}_{\mathbf{b}})=\delta
_{\mathbf{b}}^{\mathbf{a}}, \label{2}%
\end{equation}
which is called the \textit{reciprocal frame} of the frame $\{\mathbf{e}%
_{\mathbf{a}}\}$.

\textbf{5.} We define the \textit{dual frame} of the frame $\{\mathbf{e}%
_{\mathbf{a}}\}$ as the set of four covector fields (also called $1$-form
fields) $\{\varepsilon^{\mathbf{a}}\}$, $\varepsilon^{\mathbf{a}}\in\sec
T^{\ast}U\subset\sec T^{\ast}M$, $\mathbf{a}=0,1,2,3$. We also write that
$\{\varepsilon^{\mathbf{a}}\}\in\sec P_{\mathrm{SO}_{1,3}^{e}}U\subset\sec
P_{\mathrm{SO}_{1,3}^{e}}M$, i.e., it is a section of the orthonormal
\textit{coframe }bundle. \ The set $\{\varepsilon^{\mathbf{a}}\}$ is called an
orthonormal coframe.

\textbf{6.} Recall that $1$-forms are \textit{mappings} $\sec TM\rightarrow
\mathbb{R}$ and by definition the $1$-forms $\varepsilon^{\mathbf{a}}$,
$\mathbf{a}=0,1,2,3$ satisfy%
\begin{equation}
\varepsilon^{\mathbf{a}}(\mathbf{e}_{\mathbf{b}})=\delta_{\mathbf{b}%
}^{\mathbf{a}} \label{3}%
\end{equation}

\textbf{7. }If $g\in\sec T_{2}^{0}M$ is the metric of the \textit{cotangent
}bundle we have%
\begin{equation}
g(\varepsilon^{\mathbf{a}},\varepsilon^{\mathbf{b}})=\eta^{\mathbf{ab}%
}:=\mathrm{diag}(1,-1,-1,-1). \label{4}%
\end{equation}

\textbf{8.} The coframe $\{\varepsilon_{\mathbf{a}}\}\in\sec P_{\mathrm{SO}%
_{1,3}^{e}}U\subset\sec P_{\mathrm{SO}_{1,3}^{e}}M$ such that \
\begin{equation}
g(\varepsilon_{\mathbf{a}},\varepsilon^{\mathbf{b}})=\delta_{\mathbf{a}%
}^{\mathbf{b}} \label{5}%
\end{equation}
is called the reciprocal coframe of the coframe $\{\varepsilon^{\mathbf{a}}\}$.

\textbf{9.} Next introduce a coordinate chart\footnote{Recall that
$\chi:M\supset U\rightarrow\mathbb{R}^{4}$, $U\ni x\mapsto\chi(x):=(x^{0}%
(x),x^{1}(x),x^{2}(x),x^{3}(x))\in\mathbb{R}^{4}$.} $(\chi,U)$ of the maximal
atlas of $M$ ($U\subset M$) with coordinate functions \ $\{x^{\mu}\}$,
$\mu=0,1,2,3$. Then, we have the set of \textit{coordinate} vector
fields\footnote{$\partial_{\mu}:=\frac{\partial}{\partial x^{\mu}}$}
$\{\partial_{\mu}\}$, where each one of the $\partial_{\mu}\in\sec
TU\subset\sec TM$ is a basis for $TU$. We also write $\{\partial_{\mu}%
\}\in\sec\mathbf{F}U\subset\sec\mathbf{F}M$ and read that $\{\partial_{\mu}\}$
is a section of the \textit{frame bundle }$\mathbf{F}M$\textit{.}

\textbf{10. } The set of \textit{coordinate} covector fields $\{dx^{\mu}\}$,
where each one of the $dx^{\mu}\in\sec T^{\ast}U\subset\sec T^{\ast}M$ is a
basis for $T^{\ast}U$. We also write $\{dx^{\mu}\}\in\sec FU\subset\sec FM$
and read that $\{dx^{\mu}\}$ is a section of the \textit{coframe bundle
\ }$FM$ \textit{.}

\textbf{11. }Since $\{\partial_{\mu}\}$ is a basis for $TU$ we can expand any
one of the vector fields $\mathbf{e}_{\mathbf{a}}$, $\mathbf{a}=0,1,2,3$ as%
\begin{equation}
\mathbf{e}_{\mathbf{a}}=e_{\mathbf{a}}^{\mu}\partial_{\mu} \label{6}%
\end{equation}
where for each \textit{fixed} $\mathbf{a}$ the set $\{e_{\mathbf{a}}^{\mu}\}$
are the components of the vector field $\mathbf{e}_{\mathbf{a}}$ in the basis
$\{\partial_{\mu}\}$ and where for each fixed $\mathbf{a}$ and fixed $\mu$,
$e_{\mathbf{a}}^{\mu}:\mathbb{R}^{4}\supset\chi(U)\rightarrow\mathbb{R}$ ,
i.e., is a real function. Of course, we need a set of $16$ real functions to
represent the tetrad $\{\mathbf{e}_{\mathbf{a}}\}$.

\textbf{12. }\ If we denote by $\{\partial^{\mu}\}\in\sec\mathbf{F}%
U\subset\sec\mathbf{F}M$ the reciprocal frame of the frame $\{\partial_{\mu
}\}$ we can write%
\begin{equation}
\mathbf{e}^{\mathbf{a}}=e_{\mu}^{\mathbf{a}}\partial^{\mu}, \label{7}%
\end{equation}
where for each \textit{fixed} $\mathbf{a}$ the set $\{e_{\mu}^{\mathbf{a}}\}$
are the components of the vector field $\mathbf{e}^{\mathbf{a}}$ in the basis
$\{\partial^{\mu}\}$ and where for each fixed $\mathbf{a}$ and fixed $\mu$,
$e_{\mu}^{\mathbf{a}}:\mathbb{R}^{4}\supset\chi(U)\rightarrow\mathbb{R}$ ,
i.e., is a real function. Of course, we need a set of $16$ real functions to
represent the tetrad $\{\mathbf{e}^{\mathbf{a}}\}.$

\textbf{13. }Note that we write as usual
\begin{equation}
\mathbf{g}(\partial_{\mu},\partial_{\nu}\mathbf{)=}g_{\mu\nu}=g_{\nu\mu
}=\mathbf{g(\partial}_{\nu},\partial_{\mu}\mathbf{),} \label{8}%
\end{equation}
where for each fixed $\mu,\nu$, $g_{\mu\nu}:U\rightarrow\mathbb{R}$ , i.e., is
a real function. Of course there are at most $10$ independent $g_{\mu\nu}$ functions.

\textbf{14.} We have immediately that
\begin{equation}
e_{\mu}^{\mathbf{a}}e_{\mathbf{b}}^{\mu}=\delta_{\mathbf{b}}^{\mathbf{a}%
}\text{, }e_{\nu}^{\mathbf{a}}e_{\mathbf{a}}^{\mu}=\delta_{\nu}^{\mu},
\label{9}%
\end{equation}
and
\begin{equation}
g_{\mu\nu}=e_{\mu}^{\mathbf{a}}\eta_{\mathbf{ab}}e_{\nu}^{\mathbf{b}}.
\label{10}%
\end{equation}

\textbf{15. }Before proceeding it is worth to emphasize that for each $x\in
U\subset M$, any one of the vectors $\left.  \partial_{\mu}\right\vert
_{x},\left.  \mathbf{e}_{\mathbf{a}}\right\vert _{x}\in T_{x}U$, i.e., are
elements of the \textit{same} space, i.e., the tangent space $T_{x}U$, which
is the fiber over $x$ of the tangent bundle $TM$. It is \textbf{nonsense }as
written more than one time in \cite{sarfatti} to say that : " the set
$\{\left.  \partial_{\mu}\right\vert _{x}$ $\}$ is a basis of vectors in the
base curved spacetime and that the set $\{\left.  \mathbf{e}_{\mathbf{a}%
}\right\vert _{x}\}$ belongs to the tangent fiber at the same local scattering
coincidence $x\in U\subset M$."

\textbf{16. }If we denote by $\{dx_{\mu}\}\in\sec FU\subset\sec FM$ the
reciprocal coframe of the coframe $\{dx^{\mu}\}$ we can write%
\begin{equation}
\varepsilon^{\mathbf{a}}=e_{\mu}^{\mathbf{a}}dx^{\mu},\text{ }\varepsilon
_{\mathbf{a}}=e_{\mathbf{a}}^{\nu}dx_{\nu} \label{11}%
\end{equation}
where for each each \textit{fixed} $\mathbf{a}$ the set $\{e_{\mu}%
^{\mathbf{a}}\}$ (respectively $\{e_{\mathbf{a}}^{\nu}\})$ contains the
components of the covector field $\varepsilon^{\mathbf{a}}$ (respectively
$\varepsilon_{\mathbf{a}}$) in the basis $\{dx^{\mu}\}$ (respectively
$\{dx_{\nu}\}$) and where for each fixed $\mathbf{a}$ and fixed $\mu$
(respectively $\nu$), $e_{\mu}^{\mathbf{a}}:\mathbb{R}^{4}\supset
\chi(U)\rightarrow\mathbb{R}$ ($e_{\mathbf{a}}^{\nu}:\mathbb{R}^{4}\supset
\chi(U)\rightarrow\mathbb{R}$), i.e., they are real functions. Of course, we
need a set of $16$ real functions $e_{\mu}^{\mathbf{a}}$ to represent the
\textit{cotetrad} $\{\varepsilon^{\mathbf{a}}\}$. We also have
\begin{align}
g(dx^{\mu},dx^{\nu})  &  =g^{\mu\nu}=g^{\nu\mu}=g(dx^{\nu},dx^{\mu
})\nonumber\\
g(\varepsilon^{\mathbf{a}},\varepsilon^{\mathbf{b}})  &  =\eta^{\mathbf{ab}%
}:=\mathrm{diag}(1,-1,-1,-1),\nonumber\\
g(dx^{\mu},dx_{\nu})  &  =\delta_{\nu}^{\mu},\nonumber\\
g^{\mu\nu}  &  =e_{\mathbf{a}}^{\mu}\eta^{\mathbf{ab}}e_{\mathbf{b}}^{\nu
},\text{ }g^{\mu\nu}g_{\mu\alpha}=\delta_{\alpha}^{\mu}.,\nonumber\\
\mathbf{g}  &  =g_{\mu\nu}dx^{\mu}\otimes dx^{\nu}=\eta_{\mathbf{ab}%
}\mathbf{\varepsilon}^{\mathbf{a}}\otimes\mathbf{\varepsilon}^{\mathbf{b}%
},\nonumber\\
g  &  =g^{\mu\nu}\partial_{\mu}\otimes\partial_{\nu}=\eta^{\mathbf{ab}%
}\mathbf{e}_{\mathbf{a}}\otimes\mathbf{e}_{\mathbf{b}} \label{12}%
\end{align}

\textbf{17. }An observation similar to the one in \textbf{15} holds, e.g., for
anyone of the 1-forms $\left.  dx^{\mu}\right\vert _{x}$ or $\left.
\varepsilon^{\mathbf{a}}\right\vert _{x}$ which are elements of the
\textit{same space,} i.e., the \textit{cotangent} space $T_{x}^{\ast}U$, which
is the fiber over $x$ of the cotangent bundle $T^{\ast}M$.

\textbf{18.} Now, once the \textit{set} of 32 real functions $e_{\mu
}^{\mathbf{a}}$, $e_{\mathbf{b}}^{\nu}$ is known we can construct the
following tensor field
\begin{equation}
\varepsilon=e_{\mu}^{\mathbf{a}}\mathbf{e}_{\mathbf{a}}\otimes dx^{\mu
}=e_{\mathbf{a}}^{\mu}\partial_{\mu}\otimes\mathbf{\varepsilon}^{\mathbf{a}%
}\in\sec T_{1}^{1}U\subset\sec T_{1}^{1}M. \label{13}%
\end{equation}

Of course now, e.g., the set $\{e_{\mu}^{\mathbf{a}}\}$ with
$\mathbf{a=0,1,2,3}$ and $\mu=0,1,2,3$ contains the components of tensor field
of type $(1,1)$ in the hybrid basis $\{\mathbf{e}_{\mathbf{a}}\otimes dx^{\mu
}\}$ of $T_{1}^{1}U$.

Now, by definition the sections of $T_{1}^{1}U$ are mappings $\sec
TU\rightarrow\sec TU$, i.e., we have that $\varepsilon\mathbf{:}\sec
TU\rightarrow\sec TU$. Now, take an arbitrary vector field $\mathbf{v}\in\sec
TU$. Write
\begin{equation}
\mathbf{v}=v^{\alpha}\partial_{\alpha}. \label{14}%
\end{equation}
Then, using the definition of $\varepsilon$ we have
\begin{align}
\varepsilon(\mathbf{v})  &  =e_{\mu}^{\mathbf{a}}\mathbf{e}_{\mathbf{a}%
}\otimes dx^{\mu}(v^{\alpha}\partial_{\alpha})\nonumber\\
&  =e_{\mu}^{\mathbf{a}}v^{\alpha}\mathbf{e}_{\mathbf{a}}\otimes dx^{\mu
}(\partial_{\alpha})\nonumber\\
&  =e_{\mu}^{\mathbf{a}}v^{\alpha}\mathbf{e}_{\mathbf{a}}\delta_{\alpha}^{\mu
}\nonumber\\
&  =e_{\mu}^{\mathbf{a}}v^{\mu}\mathbf{e}_{\mathbf{a}}=v^{\mathbf{a}%
}\mathbf{e}_{\mathbf{a}}=\mathbf{v}. \label{15}%
\end{align}

Eq.(\ref{15}) shows that $\mathbf{\varepsilon}$ is \ nothing more than the
\textit{identity tensor }in $TU$. \ To call $\mathbf{\varepsilon}$\textbf{ the
Einstein-Cartan tetrad }$1$-form \ field seems to me a nonsense since
$\mathbf{\varepsilon}$ is a vector valued $1$-form field, and it is related to
the \ pullback of \textit{soldering form} of the theory of the linear
connections\footnote{Some presentations on these issues, like the one by
Rovelli \cite{rovelli} are very bad and adds confusion on the subject. We
discuss \ the approach involving soldering forms in the Appendix.}. For
details, please consult \cite{kono} (one of the best books I ever read on
differential geometry) or \cite{rodol} which has a more soft mathematical presentation.

\subsubsection{A Single Identity Operator Mislead as a \ `Tetrad'}

\textbf{19. }Note that we can write from Eq.(\ref{15}) that
\begin{equation}
\varepsilon=\mathbf{\delta}_{\mathbf{b}}^{\mathbf{a}}\mathbf{e}_{\mathbf{a}%
}\otimes\varepsilon^{\mathbf{b}} \label{16}%
\end{equation}

Note that there exists a chart of the maximal atlas of $M$ with coordinate
functions \textit{conveniently} denoted by $\{\xi^{\mathbf{a}}\}$ such that at
a given $x\in U\subset M$ we can take
\begin{equation}
\left.  d\xi^{\mathbf{a}}\right\vert _{x}=\left.  \varepsilon^{\mathbf{a}%
}\right\vert _{x} \label{17}%
\end{equation}

The coordinate functions $\xi^{\mathbf{a}}$, $\mathbf{a}=0,1,2,3$ are called
\textit{local Lorentz coordinates} in Physics textbooks.\footnote{In
Mathematics text books they are called Riemann normal coordinates for $U$,
based on $x\in U$.}

\textbf{20. }Of course, if we write\footnote{Note that these are equations
(1.1) and (1.2) in \cite{sarfatti}.}
\begin{align}
\varepsilon &  =e_{\mu}^{\mathbf{a}}\mathbf{e}_{\mathbf{a}}\otimes dx^{\mu
}=I+\ell\label{17bis}\\
&  =I_{\mu}^{\mathbf{a}}\mathbf{e}_{\mathbf{a}}\otimes dx^{\mu}+\ell_{\mu
}^{\mathbf{a}}\mathbf{e}_{\mathbf{a}}\otimes dx^{\mu}\label{18}\\
&  =I_{\mu}^{\prime\mathbf{a}}\mathbf{e}_{\mathbf{a}}\otimes d\xi^{\mu}%
+\ell_{\mu}^{\prime\mathbf{a}}\mathbf{e}_{\mathbf{a}}\otimes d\xi^{\mu},
\label{19}%
\end{align}
then at $x\in U$ we have that
\begin{equation}
\left.  I_{\mu}^{\prime\mathbf{a}}\right\vert _{x}=\mathrm{diag}%
(1,1,1,1)\text{, }\left.  \ell_{\mu}^{\prime\mathbf{a}}\right\vert _{x}=0.
\label{20}%
\end{equation}

\textbf{21. }Keep in mind that \ $\ell\in\sec T_{1}^{1}M$, i.e., it is a
vector valued $1$-form field.

\textbf{22. }In \cite{sarfatti} it is stated that when the "intrinsically
curved piece \ $\ell_{\mu}^{\prime\mathbf{a}}$ of $\mathbf{e}$" is null on a
region $U\subset M$ then $U$ is flat. \ Since curvature refers to a
well-defined property of a given connection defined on $M$, the above
statement has meaning only if the manifold $M$ is equipped with a given
connection, which is the \textit{Levi-Civita connection}\footnote{A covariant
derivative is a connection acting on some vector bundle associated with a
given principal bundle where a connection field is defined. Details can be
found, e.g., in \cite{kono,rodol}.} $D$ of $\mathbf{g}$. We can introduce
other general connections in $M$ such that the statement is not true.

Moreover, the converse of the statement is not true. As example, imagine that
$(M$,$\mathbf{g,}D)$ is Minkowski spacetime\footnote{More precisely,
$(M$,$\mathbf{g,}D)$ is part of the structure $(M\simeq\mathbb{R}%
^{4},\mathbf{g,}D,\tau_{\mathbf{g}},\uparrow)$ defining Minkowski spacetime.
For details, see,e.g., \cite{rodol}.}. If we introduce, e.g., spherical
coordinates in $U\subset M$ then in that coordinates $\ell_{\mu}%
^{\prime\mathbf{a}}\neq0$ on $U$. However $U$ is always flat, this last
statement meaning that the Riemann curvature tensor of $D$ is null in all
points of $M$.

\textbf{23. }A choice of a section $\{\mathbf{e}_{\mathbf{a}}\}$ of
$\mathbf{P}_{\mathrm{SO}_{1,3}^{e}}M$ will be called a \textit{choice of
gauge}. If $\{\mathbf{e}_{\mathbf{a}}^{\prime}\}$ $\in\sec$f $\mathbf{P}%
_{\mathrm{SO}_{1,3}^{e}}M$ is another choice of gauge we have
\begin{equation}
\mathbf{e}_{\mathbf{a}}^{\prime}=\mathbf{e}_{\mathbf{b}}L_{\mathbf{a}%
}^{\mathbf{b}}, \label{21}%
\end{equation}
where the matrix $\mathbf{L=}(L_{\mathbf{b}}^{\mathbf{a}}):M\supset
U\rightarrow\mathrm{SO}_{1,3}^{e}$.

\section{Connection $1$-forms}

\textbf{24. }Perhaps the most pedestrian way for introducing the connection
$1$-form \textit{fields} $\omega_{\mathbf{b}}^{\mathbf{a}}$ ($\mathbf{a,b}%
=0,1,2,3$) associated with an arbitrary metric compatible connection $\nabla$
on the manifold $M$ is the following. Introduce on $U\subset M$ coordinate
functions $\{x^{\mu}\}$ and the following bases for $TU$ and respective dual
bases for $T^{\ast}U$: $\{\partial_{\mu}\},\{\partial^{\mu}\}\in\sec
\mathbf{F}M$, $\{\mathbf{e}_{\mathbf{a}}\},\{\mathbf{e}^{\mathbf{a}}\}$
$\in\sec$ $\mathbf{P}_{\mathrm{SO}_{1,3}^{e}}M\subset\sec\mathbf{F}M$,
$\{\theta^{\mu}=dx^{\mu}\},\{\theta_{\mu}=g_{\mu\nu}dx^{\nu}\}\in\sec FM$,
$\{\varepsilon^{\mathbf{a}}\},\{\mathbf{\varepsilon}_{a}\}\in\sec$
$P_{\mathrm{SO}_{1,3}^{e}}M\subset\sec FM$.

Now define the coefficients of the connection in the various bases introduced
above by:%

\begin{align}
\nabla_{\partial_{\mu}}\partial_{\nu}  &  =\Gamma_{\mu\nu}^{\alpha}%
\partial_{\alpha}\text{, }\nabla_{\partial\sigma}\partial^{\mu}=-\Gamma
_{\sigma\alpha}^{\mu}\partial^{\alpha},\nonumber\\
\nabla_{\mathbf{e}_{\mathbf{a}}}\mathbf{e}_{\mathbf{b}}  &  =\omega
_{\mathbf{ab}}^{\mathbf{c}}\mathbf{e}_{\mathbf{c}},\qquad\nabla_{\mathbf{e}%
_{\mathbf{a}}}\mathbf{e}^{\mathbf{b}}=-\omega_{\mathbf{ac}}^{\mathbf{b}%
}\mathbf{e}^{\mathbf{c}},\;\nabla_{\mathbf{e}_{\mathbf{a}}}\varepsilon
^{\mathbf{b}}=-\omega_{\mathbf{ac}}^{\mathbf{b}}\varepsilon^{\mathbf{c}%
}\nonumber\\
\hspace{0.15cm}\nabla_{\mathbf{e}_{\mathbf{a}}}\varepsilon_{\mathbf{b}}  &
=-\omega_{\mathbf{cab}}\mathbf{\varepsilon}^{\mathbf{c}}\nonumber\\
\;\;\text{ }\omega_{\mathbf{abc}}  &  =\eta_{\mathbf{ad}}\omega_{\mathbf{bc}%
}^{\mathbf{d}}=-\omega_{\mathbf{cba}},\text{ }\omega_{\mathbf{a}}%
^{\mathbf{bc}}=\eta^{\mathbf{bk}}\omega_{\mathbf{kal}}\eta^{\mathbf{cl}%
},\text{ }\omega_{\mathbf{a}}^{\mathbf{bc}}=-\omega_{\mathbf{a}}^{\mathbf{cb}%
}\label{22}\\
\nabla_{\partial_{\mu}}\mathbf{e}_{\mathbf{b}}  &  =\omega_{\mu\mathbf{b}%
}^{\mathbf{c}}\mathbf{e}_{\mathbf{c}},\nonumber\\
\nabla_{\partial_{\mu}}dx^{\nu}  &  =-\Gamma_{\mu\alpha}^{\nu}dx^{\alpha
},\;\;\;\nabla_{\partial_{\mu}}\theta_{\nu}=\Gamma_{\mu\nu}^{\rho}\theta
_{\rho},\nonumber\\
\nabla_{\mathbf{e}_{\mathbf{a}}}\varepsilon^{\mathbf{b}}=-\omega_{\mathbf{ac}%
}^{\mathbf{b}}\varepsilon^{\mathbf{c}}  &  ,\;\;\;\nabla_{\partial_{\mu}%
}\varepsilon^{\mathbf{b}}=-\omega_{\mu\mathbf{a}}^{\mathbf{b}}\varepsilon
^{\mathbf{a}}\text{, etc}...\nonumber
\end{align}

Recall that for the Levi-Civita connection of $\mathbf{g}$ that we denote by
$D$ the connection coefficients $\Gamma_{\mu\nu}^{\alpha}$ are symmetric but
the connection \textit{coefficients} $\omega_{\mathbf{a}}^{\mathbf{bc}}$ of
the \textit{same} connection (in another basis) are antisymmetric, i.e.,
$\omega_{\mathbf{a}}^{\mathbf{bc}}=-\omega_{\mathbf{a}}^{\mathbf{cb}}$.

\textbf{25. }The covariant differential $\mathbf{\nabla}$ of a vector field
$\mathbf{v}\in\sec TM$ is the mapping:%
\begin{equation}
\mathbf{\nabla:}\sec TM\rightarrow\sec TM\otimes\sec T^{\ast}M,\text{
}\mathbf{v\rightarrow\nabla v,} \label{23}%
\end{equation}
such that for any vector field $\mathbf{X}\in\sec TM$ we have
\begin{equation}
\mathbf{\nabla v(X)=\nabla}_{\mathbf{X}}\mathbf{v.} \label{24}%
\end{equation}
We now can easily verify that the covariant differential of a basis vector
field $\mathbf{e}_{\mathbf{a}}$ is given by%
\begin{equation}
\mathbf{\nabla e}_{\mathbf{b}}=\mathbf{e}_{\mathbf{a}}\otimes\omega
_{\hspace{0.05cm}\mathbf{b}}^{\mathbf{a}}, \label{25}%
\end{equation}
where the $\omega_{\hspace{0.05cm}\mathbf{a}}^{\mathbf{b}}$ are the so called
(gauge dependent) connection $1$-form fields,%
\begin{equation}
\omega_{\hspace{0.05cm}\mathbf{b}}^{\mathbf{a}}=\omega_{\mathbf{cb}%
}^{\mathbf{a}}\mathbf{\varepsilon}^{\mathbf{c}}. \label{26}%
\end{equation}

We can immediately verify that
\begin{equation}
\omega_{\mathbf{ab}}:=\eta_{\mathbf{ac}}\omega_{\mathbf{b}}^{\mathbf{c}%
}=-\omega_{\mathbf{ba}}, \label{26bis}%
\end{equation}
a relation which is important for what follows.

\subsection{Change of Gauge}

\textbf{26. }Consider two frames $\{\mathbf{e}_{\mathbf{a}}\},\{\mathbf{e}%
_{\mathbf{a}}^{\prime}\}$ $\in\sec\mathbf{P}_{\mathrm{SO}_{1,3}^{e}}%
U\subset\sec$ $\mathbf{P}_{\mathrm{SO}_{1,3}^{e}}M\subset\sec\mathbf{F}M$
\ related as in Eq.(\ref{21}) and the respective dual coframes $\{\varepsilon
^{\mathbf{a}}\},\{\varepsilon^{\prime\mathbf{a}}\}\in\sec P_{\mathrm{SO}%
_{1,3}^{e}}U\subset\sec P_{\mathrm{SO}_{1,3}^{e}}M$. It is useful to introduce
the following matrix notation,%
\begin{align}
\mathbf{e}  &  =\mathbf{(e}_{\mathbf{0}},\mathbf{e}_{\mathbf{1}}%
,\mathbf{e}_{\mathbf{2}},\mathbf{e}_{\mathbf{3}}),\mathbf{e}^{\prime
}\mathbf{=(e}_{\mathbf{0}}^{\prime},\mathbf{e}_{\mathbf{1}}^{\prime
},\mathbf{e}_{\mathbf{2}}^{\prime},\mathbf{e}_{\mathbf{3}}^{\prime
})\nonumber\\%
\mbox{\boldmath{$\varepsilon$}}%
^{t}  &  \mathbf{=}\mathbf{(\varepsilon}^{\mathbf{0}},\varepsilon^{\mathbf{1}%
},\varepsilon^{\mathbf{2}},\varepsilon^{\mathbf{3}}\mathbf{)}^{t}\mathbf{,}%
\mbox{\boldmath{$\varepsilon$}}%
^{\prime^{t}}\mathbf{=(\varepsilon}^{\prime\mathbf{0}},\varepsilon
^{\prime\mathbf{1}},\varepsilon^{\prime\mathbf{2}},\varepsilon^{\prime
\mathbf{3}}\mathbf{)}^{t} \label{27}%
\end{align}

Under these conditions we can write Eq.(\ref{21}) as%
\begin{equation}
\mathbf{e}^{\prime}=\mathbf{eL} \label{28}%
\end{equation}

Obviously we have also%
\begin{equation}%
\mbox{\boldmath{$\varepsilon$}}%
^{\prime}=\mathbf{L}^{-1}%
\mbox{\boldmath{$\varepsilon$}}
\label{29}%
\end{equation}

\textbf{27. }Interpret\textbf{ }$\mathbb{R}^{4}$ as a vector space over the
field of real numbers with the canonical basis
\begin{equation}
\mathbf{E}^{\mathbf{0}}=(1,0,0,0)\text{, }\mathbf{E}^{\mathbf{1}%
}=(0,1,0,0)\text{, }\mathbf{E}^{\mathbf{2}}=(0,0,1,0)\text{, }\mathbf{E}%
^{\mathbf{3}}=(0,0,0,1). \label{31}%
\end{equation}

Consider another copy of (the vector space) $\mathbb{R}^{4}$, denoted here by
$^{\ast}\mathbb{R}^{4}$ with canonical basis
\begin{equation}
^{\ast}\mathbf{E}_{\mathbf{0}}=\left(
\begin{array}
[c]{c}%
1\\
0\\
0\\
0
\end{array}
\right)  \text{, }^{\ast}\mathbf{E}_{\mathbf{1}}=\left(
\begin{array}
[c]{c}%
0\\
1\\
0\\
0
\end{array}
\right)  \text{, }^{\ast}\mathbf{E}_{\mathbf{2}}=\left(
\begin{array}
[c]{c}%
0\\
0\\
1\\
0
\end{array}
\right)  \text{, }^{\ast}\mathbf{E}_{\mathbf{3}}=\left(
\begin{array}
[c]{c}%
0\\
0\\
0\\
1
\end{array}
\right)  \label{32}%
\end{equation}

Then, we can write%
\begin{equation}
\mathbf{e=e}_{\mathbf{a}}\otimes\mathbf{E}^{\mathbf{s}}\text{, }%
\mbox{\boldmath{$\varepsilon$}}%
=\varepsilon^{\mathbf{b}}\otimes^{\ast}\mathbf{E}_{\mathbf{b}} \label{33}%
\end{equation}
and say that $\mathbf{e}$ is a $\mathbb{R}^{4}$-valued vector field and $%
\mbox{\boldmath{$\varepsilon$}}%
$ is a $\mathbb{R}^{4}$-valued $1$-form field.

\textbf{28.} Recall that the set of $4\times4$ real matrices $\mathbb{R(}%
4\mathbb{)}$ is given by the tensor product $\mathbb{R}^{4}\mathbb{\otimes
}^{\ast}\mathbb{R}^{4}$, i.e., \ $\mathbb{R(}4\mathbb{)=R}^{4}\mathbb{\otimes
}^{\ast}\mathbb{R}^{4}$.\ Next we define the tensor product
\ $\mathbf{e\otimes}%
\mbox{\boldmath{$\varepsilon$}}%
$ by%
\begin{align}
\mathbf{e\otimes}%
\mbox{\boldmath{$\varepsilon$}}%
&  =\left(  \mathbf{e}_{\mathbf{a}}\otimes\mathbf{E}^{\mathbf{a}}\right)
\otimes\left(  \varepsilon^{\mathbf{b}}\otimes^{\ast}\mathbf{E}_{\mathbf{b}%
}\right)  :=\varepsilon^{\mathbf{b}}(\mathbf{e}_{\mathbf{a}})\mathbf{E}%
^{\mathbf{a}}\otimes^{\ast}\mathbf{E}_{\mathbf{b}}=\delta_{\mathbf{a}%
}^{\mathbf{b}}\mathbf{E}^{\mathbf{a}}\otimes^{\ast}\mathbf{E}_{\mathbf{b}%
}\label{34}\\
&  =\left(
\begin{array}
[c]{cccc}%
1 & 0 & 0 & 0\\
0 & 1 & 0 & 0\\
0 & 0 & 1 & 0\\
0 & 0 & 0 & 1
\end{array}
\right)  =\mathbf{I}\nonumber
\end{align}

This is usually simplified by writing the \ `product' of the matrices
$\mathbf{e}$ and $%
\mbox{\boldmath{$\varepsilon$}}%
$ as meaning:%
\begin{align}
\mathbf{e}%
\mbox{\boldmath{$\varepsilon$}}%
&  =\mathbf{(e}_{\mathbf{0}},\mathbf{e}_{\mathbf{1}},\mathbf{e}_{\mathbf{2}%
},\mathbf{e}_{\mathbf{3}})\left(
\begin{array}
[c]{c}%
\mathbf{\varepsilon}^{\mathbf{0}}\\
\varepsilon^{\mathbf{1}}\\
\varepsilon^{\mathbf{2}}\\
\varepsilon^{\mathbf{3}}%
\end{array}
\right)  =\left(
\begin{array}
[c]{cccc}%
\mathbf{\varepsilon}^{\mathbf{0}}(\mathbf{e}_{\mathbf{0}}) & 0 & 0 & 0\\
0 & \varepsilon^{\mathbf{1}}(\mathbf{e}_{\mathbf{1}}) & 0 & 0\\
0 & 0 & \varepsilon^{\mathbf{2}}(\mathbf{e}_{\mathbf{2}}) & 0\\
0 & 0 & 0 & \varepsilon^{\mathbf{3}}(\mathbf{e}_{\mathbf{3}})
\end{array}
\right) \nonumber\\
&  =\left(
\begin{array}
[c]{cccc}%
1 & 0 & 0 & 0\\
0 & 1 & 0 & 0\\
0 & 0 & 1 & 0\\
0 & 0 & 0 & 1
\end{array}
\right)  =\mathbf{I.} \label{30}%
\end{align}

\textbf{29. }Let $\{\mathbf{E}_{\mathbf{ab}}\}$ be a set of matrices which is
a basis of $\mathbb{R}(4)$. Each matrix $\mathbf{E}_{\mathbf{ab}}$ has a $1$
in line $\mathbf{a}$, column $\mathbf{b}$ and zero in their other entries.

\textbf{ }Put $\mathbf{E}_{\mathbf{a}}^{\mathbf{b}}=\eta^{\mathbf{bc}%
}\mathbf{E}_{\mathbf{ac}}$ and define the matrix \ $%
\mbox{\boldmath{$\omega$}}%
$ of $1$-form fields by
\begin{align}%
\mbox{\boldmath{$\omega$}}
&  =\omega_{\hspace{0.05cm}\mathbf{b}}^{\mathbf{a}}\otimes\mathbf{E}%
_{\mathbf{a}}^{\mathbf{b}}\nonumber\\
&  =\omega^{\mathbf{ab}}\otimes\mathbf{E}_{\mathbf{ab}} \label{35}%
\end{align}

Taking into account that $\omega^{\mathbf{ab}}=-\omega^{\mathbf{ba}}$ we can
write
\begin{align}%
\mbox{\boldmath{$\omega$}}
&  =\frac{1}{2}\omega^{\mathbf{ab}}\otimes\mathbf{E}_{\mathbf{ab}}+\frac{1}%
{2}\omega^{\mathbf{ba}}\otimes\mathbf{E}_{\mathbf{ba}}\nonumber\\
&  =\frac{1}{2}\omega^{\mathbf{ab}}\otimes(\mathbf{E}_{\mathbf{ab}}%
-\mathbf{E}_{\mathbf{ba}})\nonumber\\
&  =\frac{1}{2}\omega^{\mathbf{ab}}\otimes\mathbf{G}_{\mathbf{ab}}, \label{36}%
\end{align}
where $\mathbf{G}_{\mathbf{ab}}:=\mathbf{E}_{\mathbf{ab}}-\mathbf{E}%
_{\mathbf{ba}}$ are a set of antisymmetric matrices in $\mathbb{R(}4)$, and as
it is well known forms a basis for a representation of the Lie algebra of
\textrm{SO}$_{1,3}^{e}$ in $\mathbb{R}^{4}$.

We then can say that $%
\mbox{\boldmath{$\omega$}}%
$ is a $1$-form with values in the Lie algebra of \textrm{SO}$_{1,3}^{e}$.

\textbf{30. \ }With $\mathbf{e}$ and $%
\mbox{\boldmath{$\omega$}}%
$ defined \ as above we can write from Eq.(\ref{21}),%
\begin{equation}
\mathbf{\nabla e=e\otimes}%
\mbox{\boldmath{$\omega$}}%
, \label{37}%
\end{equation}
which we write in simplified form as
\begin{equation}
\mathbf{\nabla e}=\mathbf{e}%
\mbox{\boldmath{$\omega$}}
\label{38}%
\end{equation}

\ Since the covariant differential $\mathbf{\nabla}$ must be well defined,
i.e., independent of the basis used, we must have
\begin{equation}
\mathbf{\nabla(e}^{\prime}\mathbf{)=e}^{\prime}%
\mbox{\boldmath{$\omega$}}%
^{\prime}=\mathbf{\nabla(eL)=(\nabla e)L+e}d\mathbf{L,} \label{39}%
\end{equation}
or,
\begin{equation}
\mathbf{eL}^{\mathbf{\ }}%
\mbox{\boldmath{$\omega$}}%
^{\prime}=\mathbf{e}%
\mbox{\boldmath{$\omega$}}%
\mathbf{L+e}d\mathbf{L,} \label{40}%
\end{equation}
from were we immediately get%
\begin{equation}
^{\mathbf{\ }}%
\mbox{\boldmath{$\omega$}}%
^{\prime}=\mathbf{L}^{-1}%
\mbox{\boldmath{$\omega$}}%
\mathbf{L+L}^{-1}d\mathbf{L.} \label{41}%
\end{equation}

\textbf{31.} \textbf{ }In \cite{sarfatti} the connection $1$-form fields are
denoted by $\Sigma^{\mathbf{ab}\text{ }}$instead of $\omega^{\mathbf{ab}}$.
Sarfatti calls the $\Sigma^{\mathbf{ab}\text{ }}$Sarfatti the
\textquotedblleft spin connection $1$-form\textquotedblright. This wording is
misleading because it did not leave clear that $\{\Sigma^{\mathbf{ab}\text{ }%
}\}$ refers to a set of \textit{six different }$1$-form fields.\textit{ }
Having introduced the set $\{\Sigma^{\mathbf{ab}\text{ \ }}\}$ Sarfatti
presents in his Eq.(1.9) an object called $\mathbf{\Sigma}$, written as:
\begin{equation}%
\begin{tabular}
[c]{|c|}\hline
$\mathbf{\Sigma=\Sigma}_{\mu}^{\mathbf{ab}}dx^{\mu}\partial_{\mathbf{a}%
}\partial_{\mathbf{b}}$\\\hline
\end{tabular}
\ \ \ \ \ \ \ \ \ \tag{1.9}%
\end{equation}
and call it the \ \textquotedblleft local invariant spin-connection $1$-form
in curved spacetime\textquotedblright. \ This denomination containing the word
\textit{invariant} is misleading, since $\mathbf{\Sigma}$ is \textit{not }a
tensor field. There is a different $\mathbf{\Sigma}$ for each choice of gauge
as it is clear from Eq.(\ref{41}) above. If the $\partial_{\mathbf{a}}$ is
interpreted as meaning the vector fields\footnote{Please, do not call the
$\partial_{\mathbf{a}}$ \textit{coforms}...} $\mathbf{e}_{\mathbf{a}}$ as
introduced above and if we introduce the Clifford bundle $\mathcal{C\ell
(}M,\mathbf{g})$ of multivector fields we can write as in \cite{olro,rodol}
\begin{align}
\mathbf{\Sigma}  &  =\mathbf{\Sigma}_{\mu}^{\mathbf{ab}}dx^{\mu}%
\otimes\mathbf{e}_{\mathbf{a}}\otimes\mathbf{e}_{\mathbf{b}}\nonumber\\
&  =\frac{1}{2}\mathbf{\Sigma}_{\mu}^{\mathbf{ab}}dx^{\mu}\otimes
\mathbf{e}_{\mathbf{a}}\wedge\mathbf{e}_{\mathbf{b}}=\frac{1}{2}%
\mathbf{\Sigma}_{\mu}^{\mathbf{ab}}dx^{\mu}\otimes\mathbf{e}_{\mathbf{a}%
}\mathbf{e}_{\mathbf{b}}, \label{42}%
\end{align}
where $\mathbf{e}_{\mathbf{a}}\mathbf{e}_{\mathbf{b}}=\mathbf{g(e}%
_{\mathbf{a}},\mathbf{e}_{\mathbf{b}})+\mathbf{e}_{\mathbf{b}}\wedge
\mathbf{e}_{\mathbf{a}}$ is the Clifford product of the vector fields
$\mathbf{e}_{\mathbf{a}}$ and $\mathbf{e}_{\mathbf{b}}$ interpreted as
sections of $\mathcal{C\ell(}M,\mathbf{g})$. As it is well known (see, e.g.,
\cite{rodol}) the bivector fields $\mathbf{e}_{\mathbf{ab}}=\mathbf{e}%
_{\mathbf{a}}\mathbf{e}_{\mathbf{b}}$ generate the Lie algebra of
\textrm{SO}$_{1,3}^{e}\simeq\mathrm{Sl}(2,\mathbb{C)\simeq}\mathrm{Spin}%
_{1,3}^{e}$ and so we arrive again at another (equivalent) description of the
connection $1$-forms fields.

\section{Vector-valued $r$-forms,Torsion and Curvature}

\textbf{32. }In what follows we denote by $%
\mbox{\boldmath{$\tau$}}%
M=%
{\displaystyle\sum\nolimits_{r,s=0}^{\infty}}
\oplus T_{s}^{r}M$ the \textit{tensor bundle} of $M$ \ and by $%
{\displaystyle\bigwedge}
T^{\ast}M=%
{\displaystyle\sum\nolimits_{r=0}^{4}}
\oplus%
{\displaystyle\bigwedge\nolimits^{r}}
T^{\ast}M$ the \textit{exterior bundle} of $M$. Note that$%
{\displaystyle\bigwedge\nolimits^{r}}
T^{\ast}M$ is the bundle of $r$-forms and we have the identification $%
{\displaystyle\bigwedge\nolimits^{1}}
T^{\ast}M=T^{\ast}M$ and $%
{\displaystyle\bigwedge\nolimits^{0}}
T^{\ast}M=\mathcal{F(}M)$ the set of differentiable functions on $M$. Also, by
$%
{\displaystyle\bigwedge}
TM=%
{\displaystyle\sum\nolimits_{r=0}^{4}}
\oplus%
{\displaystyle\bigwedge\nolimits^{r}}
TM$ $\ $we denote the exterior bundle of multivectors. We have also the
identifications $%
{\displaystyle\bigwedge\nolimits^{1}}
TM=TM$ and $%
{\displaystyle\bigwedge\nolimits^{0}}
TM=\mathcal{F(}M)$.

A vector valued $r$-form $%
\mbox{\boldmath{$\alpha$}}%
$\ is a section of the bundle $TM\otimes%
{\displaystyle\bigwedge\nolimits^{r}}
T^{\ast}M$ \ which we write using the basis $\{\mathbf{e}_{\mathbf{a}}\}$ of
$TM$ as%
\begin{equation}%
\mbox{\boldmath{$\alpha$}}%
=\mathbf{e}_{\mathbf{a}}\otimes\alpha^{\mathbf{a}}, \label{43}%
\end{equation}
where $\alpha^{\mathbf{a}}\in\sec%
{\displaystyle\bigwedge\nolimits^{r}}
T^{\ast}M$.

\textbf{33. }We introduce next the exterior covariant differential of a vector
valued $r$-form $%
\mbox{\boldmath{$\alpha$}}%
$ as the $(r+1)$-form $\mathbf{\nabla}%
\mbox{\boldmath{$\alpha$}}%
$ such that
\begin{equation}
\mathbf{\nabla}%
\mbox{\boldmath{$\alpha$}}%
=\mathbf{\nabla}(\mathbf{e}_{\mathbf{a}}\otimes\alpha^{\mathbf{a}%
}):=(\mathbf{\nabla e}_{\mathbf{a}})\otimes_{\wedge}\alpha^{\mathbf{a}%
}+\mathbf{e}_{\mathbf{a}}\otimes d\alpha^{\mathbf{a}}. \label{44}%
\end{equation}

The product $\otimes_{\wedge}$is defined by%
\begin{equation}
(\mathbf{\nabla e}_{\mathbf{b}})\otimes_{\wedge}\alpha^{\mathbf{b}%
}=(\mathbf{e}_{\mathbf{a}}\otimes\omega_{\mathbf{b}}^{\mathbf{a}}%
)\otimes_{\wedge}\alpha^{\mathbf{b}}:=\mathbf{e}_{\mathbf{a}}\otimes
(\omega_{\mathbf{b}}^{\mathbf{a}}\wedge\alpha^{\mathbf{b}}). \label{45}%
\end{equation}

Then, from Eq.(\ref{44}) we get%
\begin{equation}
\mathbf{\nabla}%
\mbox{\boldmath{$\alpha$}}%
=\mathbf{e}_{\mathbf{a}}\otimes(d\alpha^{\mathbf{a}}+\omega_{\mathbf{b}%
}^{\mathbf{a}}\wedge\alpha^{\mathbf{b}}) \label{46}%
\end{equation}

\textbf{34. }Before continuing we observe that we denoted the exterior
covariant differential \ by the same symbol as the covariant differential
because it is in a sense an extension of this later object which was has been
introduced\ above as a mapping (satisfying certain properties) sending \ $\sec
TM\rightarrow\sec TM\otimes\sec%
{\displaystyle\bigwedge\nolimits^{1}}
T^{\ast}M$ .

\textbf{35. }\ Observe that $%
\mbox{\boldmath{$\varepsilon$}}%
=\mathbf{e}_{\mathbf{a}}\otimes\varepsilon^{\mathbf{a}}$ is a vector valued
$1$-form. Then, its exterior covariant differential is:%
\begin{equation}
\mathbf{\nabla}%
\mbox{\boldmath{$\varepsilon$}}%
=\mathbf{e}_{\mathbf{a}}\otimes(d\varepsilon^{\mathbf{a}}+\omega_{\mathbf{b}%
}^{\mathbf{a}}\wedge\varepsilon^{\mathbf{b}}) \label{47}%
\end{equation}

Now, recalling that
\begin{align}
d\varepsilon^{\mathbf{a}}(\mathbf{e}_{\mathbf{b}},\mathbf{e}_{\mathbf{c}})  &
=\mathbf{e}_{\mathbf{b}}(\varepsilon^{\mathbf{a}}(\mathbf{e}_{\mathbf{c}%
}))-\mathbf{e}_{\mathbf{c}}(\varepsilon^{\mathbf{a}}(\mathbf{e}_{\mathbf{b}%
}))-\varepsilon^{\mathbf{a}}(\left[  \mathbf{e}_{\mathbf{b}},\mathbf{e}%
_{\mathbf{c}}\right]  )\nonumber\\
&  =-\varepsilon^{\mathbf{a}}(\left[  \mathbf{e}_{\mathbf{b}},\mathbf{e}%
_{\mathbf{c}}\right]  )=-\varepsilon^{\mathbf{a}}\left(  \nabla_{\mathbf{e}%
_{\mathbf{b}}}\mathbf{e}_{\mathbf{c}}-\nabla_{\mathbf{e}_{\mathbf{c}}%
}\mathbf{e}_{\mathbf{b}}-\tau(\mathbf{e}_{\mathbf{b}},\mathbf{e}_{\mathbf{c}%
})\right) \label{48}\\
&  =-\varepsilon^{\mathbf{a}}\left(  \omega_{\mathbf{bc}}^{\mathbf{d}%
}\mathbf{e}_{\mathbf{d}}-\omega_{\mathbf{cb}}^{\mathbf{d}}\mathbf{e}%
_{\mathbf{d}}-\tau(\mathbf{e}_{\mathbf{b}},\mathbf{e}_{\mathbf{c}})\right) \\
&  =-(\omega_{\mathbf{bc}}^{\mathbf{a}}-\omega_{\mathbf{cb}}^{\mathbf{a}%
})+T_{\mathbf{bc}}^{\mathbf{a}}.
\end{align}
(where we used \ that $\tau\in\sec TM\otimes\sec%
{\displaystyle\bigwedge\nolimits^{2}}
T^{\ast}M$) the vector valued torsion form is given by%
\begin{equation}
\tau=\mathbf{e}_{\mathbf{a}}\otimes\tau^{\mathbf{a}}=\frac{1}{2}%
\mathbf{e}_{\mathbf{a}}\otimes T_{\mathbf{bc}}^{\mathbf{a}}\varepsilon
^{\mathbf{b}}\wedge\varepsilon^{\mathbf{c}}, \label{50}%
\end{equation}
where the $\tau^{\mathbf{a}}=\frac{1}{2}T_{\mathbf{bc}}^{\mathbf{a}%
}\varepsilon^{\mathbf{b}}\wedge\varepsilon^{\mathbf{c}}$ $\in\sec%
{\displaystyle\bigwedge\nolimits^{2}}
T^{\ast}M$ are called the \textit{torsion} $2$-forms.

For eventual future reference we also recall that%
\begin{equation}
\left[  \mathbf{e}_{\mathbf{b}},\mathbf{e}_{\mathbf{c}}\right]
=c_{\mathbf{bc}}^{\mathbf{d}}\mathbf{e}_{\mathbf{d}}, \label{49}%
\end{equation}
and that
\begin{equation}
d\varepsilon^{\mathbf{a}}:=-\frac{1}{2}c_{\mathbf{bc}}^{\mathbf{a}}%
\varepsilon^{\mathbf{b}}\wedge\varepsilon^{\mathbf{c}}, \label{51}%
\end{equation}
from where we get
\begin{equation}
T_{\mathbf{bc}}^{\mathbf{a}}=\omega_{\mathbf{bc}}^{\mathbf{a}}-\omega
_{\mathbf{cb}}^{\mathbf{a}}-c_{\mathbf{bc}}^{\mathbf{a}}. \label{52}%
\end{equation}

\textbf{36. \ }We then have,%
\begin{equation}
\tau^{\mathbf{a}}=d\varepsilon^{\mathbf{a}}+\omega_{\mathbf{b}}^{\mathbf{a}%
}\wedge\varepsilon^{\mathbf{b}}, \label{53}%
\end{equation}
known as Cartan's \textit{first} structure equation.

\textbf{37.} Defining
\begin{equation}%
\mbox{\boldmath{$\tau$}}%
^{t}=(\tau^{\mathbf{0}},\tau^{\mathbf{1}},\tau^{\mathbf{2}},\tau^{\mathbf{3}})
\label{54}%
\end{equation}
we may write (in obvious notation)
\begin{equation}%
\mbox{\boldmath{$\tau$}}%
=d%
\mbox{\boldmath{$\varepsilon$}}%
+%
\mbox{\boldmath{$\omega$}}%
\wedge%
\mbox{\boldmath{$\varepsilon$}}%
. \label{55}%
\end{equation}

\textbf{38.} Observe that since $\mathbf{\nabla e}_{\mathbf{b}}\in\sec
TM\otimes%
{\displaystyle\bigwedge\nolimits^{1}}
T^{\ast}M$ then $\mathbf{\nabla(\nabla e}_{\mathbf{b}})\sec TM\otimes%
{\displaystyle\bigwedge\nolimits^{2}}
T^{\ast}M$ we have%
\begin{align}
\mathbf{\nabla(\nabla e}_{\mathbf{b}})  &  =\mathbf{\nabla(e}_{\mathbf{a}%
}\otimes\omega_{\mathbf{b}}^{\mathbf{a}})=\mathbf{(\nabla e}_{\mathbf{a}%
})\otimes_{\wedge}\omega_{\mathbf{b}}^{\mathbf{a}}+\mathbf{e}_{\mathbf{a}%
}\otimes d\omega_{\mathbf{b}}^{\mathbf{a}}\nonumber\\
&  =(\mathbf{e}_{\mathbf{c}}\otimes\omega_{\mathbf{a}}^{\mathbf{c}}%
)\otimes_{\wedge}\omega_{\mathbf{b}}^{\mathbf{a}}+\mathbf{e}_{\mathbf{a}%
}\otimes d\omega_{\mathbf{b}}^{\mathbf{a}}\nonumber\\
&  =\mathbf{e}_{\mathbf{a}}\otimes(\omega_{\mathbf{c}}^{\mathbf{a}}%
\wedge\omega_{\mathbf{b}}^{\mathbf{c}})+\mathbf{e}_{\mathbf{a}}\otimes
d\omega_{\mathbf{b}}^{\mathbf{a}}\nonumber\\
&  =\mathbf{e}_{\mathbf{a}}\otimes(d\omega_{\mathbf{b}}^{\mathbf{a}}%
+\omega_{\mathbf{c}}^{\mathbf{a}}\wedge\omega_{\mathbf{b}}^{\mathbf{c}%
})\nonumber\\
&  =:\mathbf{e}_{\mathbf{a}}\otimes\mathcal{R}_{\mathbf{b}}^{\mathbf{a}},
\label{56}%
\end{align}
where the $\mathcal{R}_{\mathbf{b}}^{\mathbf{a}}\in\sec%
{\displaystyle\bigwedge\nolimits^{2}}
T^{\ast}M$ are called \textit{curvature} $2$-forms. The equations which define
$\mathcal{R}_{\mathbf{b}}^{\mathbf{a}}$ ($\mathbf{a,b=0,1,2,3}$), i.e.,
\begin{equation}
\mathcal{R}_{\mathbf{b}}^{\mathbf{a}}=d\omega_{\mathbf{b}}^{\mathbf{a}}%
+\omega_{\mathbf{c}}^{\mathbf{a}}\wedge\omega_{\mathbf{b}}^{\mathbf{c}},
\label{57}%
\end{equation}
is known as Cartan's second structure equation.

\textbf{39. }With the above \ `technology' it is now an easy task to show
that
\begin{equation}
\mathcal{R}_{\mathbf{b}}^{\mathbf{a}}=\frac{1}{2}R_{\mathbf{bcd}}^{\mathbf{a}%
}\varepsilon^{\mathbf{c}}\wedge\varepsilon^{\mathbf{d}}, \label{58}%
\end{equation}
where $R_{\mathbf{bcd}}^{\mathbf{a}}$ are the components of the Riemann
curvature tensor in the "orthonormal frame" $\{\mathbf{e}_{\mathbf{a}}%
\otimes\varepsilon^{\mathbf{b}}\otimes\varepsilon^{\mathbf{c}}\otimes
\varepsilon^{\mathbf{d}}\}$ of $T_{1}^{3}U\subset T_{1}^{3}M$.

\textbf{40. }Let us calculate $\mathbf{\nabla\nabla\varepsilon=\nabla
(\nabla(e}_{\mathbf{b}}\otimes\varepsilon^{\mathbf{b}}))$. We have:%
\begin{align}
\mathbf{\nabla(\nabla(e}_{\mathbf{b}}\otimes\varepsilon^{\mathbf{b}}))  &
=\mathbf{\nabla\lbrack(\nabla e}_{\mathbf{b}})\otimes_{\wedge}\varepsilon
^{\mathbf{b}}+\mathbf{e}_{\mathbf{b}}\otimes d\varepsilon^{\mathbf{b}%
}]\nonumber\\
&  =\mathbf{\nabla}\left[  \mathbf{e}_{\mathbf{a}}\otimes\left(
d\varepsilon^{\mathbf{a}}+\omega_{\mathbf{b}}^{\mathbf{a}}\wedge
\varepsilon^{\mathbf{b}}\right)  \right] \nonumber\\
&  =(\mathbf{\nabla e}_{\mathbf{a}})\otimes_{\wedge}\left(  d\varepsilon
^{\mathbf{a}}+\omega_{\mathbf{b}}^{\mathbf{a}}\wedge\varepsilon^{\mathbf{b}%
}\right)  +\mathbf{e}_{\mathbf{a}}\otimes d\left[  d\varepsilon^{\mathbf{a}%
}+\omega_{\mathbf{b}}^{\mathbf{a}}\wedge\varepsilon^{\mathbf{b}}\right]
\nonumber\\
&  =\mathbf{e}_{\mathbf{a}}\otimes\omega_{\mathbf{c}}^{\mathbf{a}}%
\wedge(d\varepsilon^{\mathbf{c}}+\omega_{\mathbf{b}}^{\mathbf{c}}%
\wedge\varepsilon^{\mathbf{b}})+\mathbf{e}_{\mathbf{a}}\otimes\left[
d\omega_{\mathbf{b}}^{\mathbf{a}}\wedge\varepsilon^{\mathbf{b}}-\omega
_{\mathbf{b}}^{\mathbf{a}}\wedge d\varepsilon^{\mathbf{b}}\right] \nonumber\\
&  =\mathbf{e}_{\mathbf{a}}\otimes(\omega_{\mathbf{c}}^{\mathbf{a}}%
\wedge\omega_{\mathbf{b}}^{\mathbf{c}}+d\omega_{\mathbf{b}}^{\mathbf{a}%
})\wedge\varepsilon^{\mathbf{b}}\label{59}\\
&  =\mathbf{e}_{\mathbf{a}}\otimes\mathcal{R}_{\mathbf{b}}^{\mathbf{a}}%
\wedge\varepsilon^{\mathbf{b}}.\nonumber
\end{align}

\textbf{41. }Also, a calculation of $\mathbf{\nabla\nabla e}$ gives (in
obvious notation):%
\begin{equation}
\mathbf{\nabla\nabla e=e(}d%
\mbox{\boldmath{$\omega$}}%
+%
\mbox{\boldmath{$\omega$}}%
\wedge%
\mbox{\boldmath{$\omega$}}%
\mathbf{)} \label{60}%
\end{equation}

\section{$(p+q)$-indexed $\mathbf{r}$-forms}

\textbf{42. }We already meet some indexed $r$-forms, namely the torsion
$2$-forms \ $\tau^{\mathbf{a}}$ and the curvature $2$-forms $\mathcal{R}%
_{\mathbf{b}}^{\mathbf{a}}$. A general $(p+q$)-indexed $r$-form is an object
defined as follows. Suppose that $X\in\sec T_{q}^{r+p}M$ and let
\begin{equation}
X_{\mathbf{\nu}_{1}....\mathbf{\nu}_{q}}^{\mathbf{\mu}_{1}....\mathbf{\mu}%
_{p}}(\mathbf{e}_{1},...,\mathbf{e}_{r})\in\sec\bigwedge\nolimits^{r}T^{\ast
}M, \label{559new}%
\end{equation}
such that
\begin{equation}
X_{\mathbf{\nu}_{1}....\mathbf{\nu}_{q}}^{\mathbf{\mu}_{1}....\mathbf{\mu}%
_{p}}(\mathbf{e}_{1},...,\mathbf{e}_{r})=X(\mathbf{e}_{1},...,\mathbf{e}%
_{r},\mathbf{e}_{\nu_{1}},....\mathbf{e}_{\nu_{q}},\varepsilon^{\mathbf{\mu
}_{1}},...,\varepsilon^{\mathbf{\mu}_{p}}). \label{559new1}%
\end{equation}

\textbf{43. }The exterior covariant derivative (differential) $\mathbf{D}%
$\textbf{ }of a $(p+q$)-indexed $r$-form $X_{\nu_{1}....\nu_{q}}^{\mu
_{1}....\mu_{p}}$ on a manifold with a general connection $\nabla$ is the mapping:%

\begin{equation}
\mathbf{D:}\sec\bigwedge\nolimits^{r}T^{\ast}M\rightarrow\sec\bigwedge
\nolimits^{r+1}T^{\ast}M\text{, }0\leq r\leq4, \label{559new2}%
\end{equation}
such that\footnote{As usual the inverted hat over a symbol (in
Eq.(\ref{559new3})) means that the corresponding symbol is missing in the
expression.}%
\begin{align}
&  (r+1)\mathbf{D}X_{\nu_{1}....\nu_{q}}^{\mathbf{\mu}_{1}....\mathbf{\mu}%
_{p}}(\mathbf{e}_{0},\mathbf{e}_{1},...,\mathbf{e}_{r})\nonumber\\
&  =\sum\limits_{\nu=0}^{r}(-1)^{\nu}\nabla_{\mathbf{e}_{\nu}}X(\mathbf{e}%
_{0},\mathbf{e}_{1},...,\mathbf{\check{e}}_{\nu},...\mathbf{e}_{r}%
,\mathbf{e}_{\nu_{1}},....\mathbf{e}_{\nu_{q}},\varepsilon^{\mathbf{\mu}_{1}%
},...,\varepsilon^{\mathbf{\mu}_{p}})\nonumber\\
&  -\sum\limits_{0\leq\nu,\varsigma\,\leq r}(-1)^{\nu+\varsigma}%
X(\mathbf{T(e}_{\nu},\mathbf{e}_{\varsigma}),\mathbf{e}_{0},\mathbf{e}%
_{1},...,\mathbf{\check{e}}_{\nu},...,\mathbf{\check{e}}_{\varsigma
},...,\mathbf{e}_{r},\mathbf{e}_{\nu_{1}},....\mathbf{e}_{\nu_{q}}%
,\varepsilon^{\mathbf{\mu}_{1}},...,\varepsilon^{\mathbf{\mu}_{p}}).
\label{559new3}%
\end{align}

Then, we may verify that
\begin{align}
\mathbf{D}X_{\mathbf{\nu}_{1}....\mathbf{\nu}_{q}}^{\mathbf{\mu}%
_{1}....\mathbf{\mu}_{p}}  &  =dX_{\mathbf{\nu}_{1}....\mathbf{\nu}_{q}%
}^{\mathbf{\mu}_{1}....\mathbf{\mu}_{p}}+\omega_{\mathbf{\mu}_{s}%
}^{\mathbf{\mu}_{1}}\wedge X_{\nu_{1}....\nu_{q}}^{\mathbf{\mu}_{s}%
....\mathbf{\mu}_{p}}+...+\omega_{\mathbf{\mu}_{s}}^{\mathbf{\mu}_{1}}\wedge
X_{\mathbf{\nu}_{1}....\mathbf{\nu}_{q}}^{\mathbf{\mu}_{1}....\mathbf{\mu}%
_{p}}\label{559new4}\\
&  +\omega_{\mathbf{\nu}_{1}}^{\nu_{s}}\wedge X_{\nu_{s}....\nu_{q}%
}^{\mathbf{\mu}_{1}....\mathbf{\mu}_{p}}+...+\omega_{\mathbf{\mu}_{s}%
}^{\mathbf{\mu}_{1}}\wedge X_{\mathbf{\nu}_{1}....\nu_{s}}^{\mathbf{\mu}%
_{1}....\mathbf{\mu}_{p}}.\nonumber
\end{align}

\textbf{44.} Note that if \emph{Eq.(\ref{559new4})} is applied on the
connection $1$-forms $\omega_{\mathbf{b}}^{\mathbf{a}}$ we would get
$\mathbf{D}\omega_{\mathbf{b}}^{\mathbf{a}}=d\omega_{\mathbf{b}}^{\mathbf{a}%
}+\omega_{\mathbf{c}}^{\mathbf{a}}\wedge\omega_{\mathbf{b}}^{\mathbf{c}%
}-\omega_{\mathbf{b}}^{\mathbf{c}}\wedge\omega_{\mathbf{c}}^{\mathbf{a}}$. So,
we see that the equation
\begin{equation}
\mathbf{D}\omega_{\mathbf{b}}^{\mathbf{a}}=\mathcal{R}_{\mathbf{b}%
}^{\mathbf{a}} \label{X}%
\end{equation}
which appears in many textbooks, and in particular as Eq.(1.11) in
\cite{sarfatti} with the substitutions $\omega_{\mathbf{b}}^{\mathbf{a}%
}\mapsto\Sigma_{\mathbf{b}}^{\mathbf{a}}$ and $\mathcal{R}_{\mathbf{b}%
}^{\mathbf{a}}\mapsto R_{\mathbf{b}}^{\mathbf{a}}$ is meaningless. The reason
for many authors to write an equation like Eq.(\ref{X}) is the wish to have an
equation similar to one that appears in the fiber bundle theory formulation of
the theory of connections. We are not going to recall that theory here. An
interested reader may study, e.g., \cite{kono,rodol}.

\section{The Einstein-Hilbert Lagrangian Density}

\textbf{45}. We recall that the Hodge star operator is the mapping%
\begin{equation}
\star:\sec%
{\displaystyle\bigwedge\nolimits^{p}}
T^{\ast}M\rightarrow\sec%
{\displaystyle\bigwedge\nolimits^{n-p}}
T^{\ast}M, \label{eh1}%
\end{equation}
such that for any $A,$ $B\in\sec%
{\displaystyle\bigwedge\nolimits^{p}}
T^{\ast}M$%
\begin{equation}
A\wedge\star B=(A\cdot B)\tau_{\mathbf{g}}, \label{eh2}%
\end{equation}
where $A\cdot B$ is the scalar product of $p$-forms\footnote{If $A=u_{1}%
\wedge...\wedge u_{p}$ and $B=v_{1}\wedge...\wedge v_{p}$, $u_{i},v_{j}\in\sec%
{\displaystyle\bigwedge\nolimits^{1}}
T^{\ast}M$, then $A\cdot B=\det g(u_{i},v_{j})$. See details in
e.g.,\cite{rodol}.} and $\tau_{\mathbf{g}}\in\sec%
{\displaystyle\bigwedge\nolimits^{4}}
T^{\ast}M$ is the volume $4$-form, which can be written with the previously
introduced notations as%
\begin{equation}
\tau_{\mathbf{g}}=\sqrt{-\det\mathbf{g}}dx^{0}\wedge dx^{1}\wedge dx^{2}\wedge
dx^{3}=\varepsilon^{\mathbf{0}}\wedge\varepsilon^{\mathbf{1}}\wedge
\varepsilon^{\mathbf{2}}\wedge\varepsilon^{\mathbf{3}}. \label{eh3}%
\end{equation}
It is useful to recall that%
\begin{equation}
\star(\varepsilon^{\mathbf{i}_{1}}\wedge...\varepsilon^{\mathbf{i}_{p}}%
)=\frac{1}{(4-p)!}\eta^{\mathbf{i}_{1}\mathbf{j}_{1}}...\eta^{\mathbf{i}%
_{p}\mathbf{j}_{p}}\epsilon_{\mathbf{j}_{1}...\mathbf{j}_{4}}\varepsilon
^{\mathbf{j}_{p+1}}\wedge...\wedge\varepsilon^{\mathbf{j}_{4}}, \label{eh3bis}%
\end{equation}
where $\varepsilon_{\mathbf{0123}}=1$ and $\epsilon_{\mathbf{j}_{1}%
...\mathbf{j}_{4}}=1$ if $\mathbf{j}_{1}...\mathbf{j}_{4}$ is an even
permutation of $(0123)$, $\epsilon_{\mathbf{j}_{1}...\mathbf{j}_{4}}=-1$ if
$\mathbf{j}_{1}...\mathbf{j}_{4}$ is an odd permutation of $(0123)$ and
$\epsilon_{\mathbf{j}_{1}...\mathbf{j}_{4}}=0$ if there are two equal digits
in the string $\mathbf{j}_{1}...\mathbf{j}_{4}$.

\textbf{46. }The Einstein-Hilbert action (in geometrical units) with
cosmological constant is%
\begin{equation}
S_{gravity}=\frac{1}{2}%
{\displaystyle\int}
(R+\Lambda)\tau_{\mathbf{g}}=\int\mathcal{L}_{gravity} \label{eh4}%
\end{equation}
where $R\in\sec%
{\displaystyle\bigwedge\nolimits^{0}}
T^{\ast}M$ is the scalar curvature \ and $\Lambda$ is a constant called the
cosmological constant \ and $\mathcal{L}_{gravity}\in\sec%
{\displaystyle\bigwedge\nolimits^{4}}
T^{\ast}M$ is the \textit{Lagrangian density} also called \textit{density of
action}.

Note that we can write $\mathcal{L}_{gravity}$ in very different but
equivalent forms, one of them very convenient for the application of the
variational formalism. It is:
\begin{equation}
\mathcal{L}_{gravity}=\frac{1}{2}\mathcal{R}_{\mathbf{cd}}\wedge
\star(\varepsilon^{\mathbf{c}}\wedge\varepsilon^{\mathbf{d}})+\frac{1}%
{2}\Lambda\tau_{\mathbf{g}}=\frac{1}{2}\star\mathcal{R}_{\mathbf{cd}}%
\wedge(\varepsilon^{\mathbf{c}}\wedge\varepsilon^{\mathbf{d}})+\frac{1}%
{2}\Lambda\tau_{\mathbf{g}} \label{eh5}%
\end{equation}

Note also that we can write correctly%
\begin{equation}
\mathcal{L}_{gravity}=\frac{1}{4}\epsilon_{\mathbf{abcd}}(\mathcal{R}%
^{\mathbf{ab}}\wedge\varepsilon^{\mathbf{c}}\wedge\varepsilon^{\mathbf{d}%
}+\frac{1}{12}\Lambda\varepsilon^{\mathbf{a}}\wedge\varepsilon^{\mathbf{b}%
}\wedge\varepsilon^{\mathbf{c}}\wedge\varepsilon^{\mathbf{d}}) \label{eh6}%
\end{equation}

\textbf{47. }Now, we may comment that Eq.(1.13) of \cite{sarfatti} has no
mathematical meaning at all, since besides representing $\mathcal{L}%
_{gravity}$ by "$\frac{%
\mbox{\boldmath{$\delta$}}%
S_{gravity}}{%
\mbox{\boldmath{$\delta$}}%
x^{4}}$" , a nonsequitur, it is written that%
\begin{equation}%
\begin{tabular}
[c]{|c|}\hline
$\frac{%
\mbox{\boldmath{$\delta$}}%
S_{gravity}}{%
\mbox{\boldmath{$\delta$}}%
x^{4}}=\ast(\mathcal{R}^{\mathbf{ab}}\wedge\varepsilon^{\mathbf{c}}%
\wedge\varepsilon^{\mathbf{d}}+\Lambda\varepsilon^{\mathbf{a}}\wedge
\varepsilon^{\mathbf{b}}\wedge\varepsilon^{\mathbf{c}}\wedge\varepsilon
^{\mathbf{d}}),$\\\hline
\end{tabular}
\ \ \tag{1.13S}%
\end{equation}
where the symbol $\ast$ is defined in a misleading and incorrect way.

We introduce the complete Lagrangian density as
\begin{equation}
\mathcal{L}\mathcal{=L}_{gravity}+\mathcal{L}_{matter}. \label{eh7}%
\end{equation}
Now, we have%
\begin{align*}%
\mbox{\boldmath{$\delta$}}%
\mathcal{L}_{matter}  &  :=%
\mbox{\boldmath{$\delta$}}%
\varepsilon^{\mathbf{a}}\wedge\frac{\partial\mathcal{L}_{matter}}%
{\partial\varepsilon^{\mathbf{a}}}\\
&  =\frac{1}{2}%
\mbox{\boldmath{$\delta$}}%
\varepsilon^{\mathbf{a}}\wedge\mathcal{T}_{\mathbf{a}}%
\end{align*}
where $\mathcal{T}_{\mathbf{a}}=T_{\mathbf{a}}^{\mathbf{b}}\varepsilon
_{\mathbf{b}}\in\sec%
{\displaystyle\bigwedge\nolimits^{1}}
T^{\ast}M$, $\mathbf{a,b=0,1,2,3}$ are the energy-momentum $1$-form fields,
such that $\mathbf{T=}$ $T_{\mathbf{a}}^{\mathbf{b}}\mathbf{e}^{\mathbf{a}%
}\otimes\varepsilon_{\mathbf{b}}\in\sec T_{1}^{1}M$ is the energy-momentum
tensor of matter.

Variation of the total action gives Einstein equations (see details, e.g., in
\cite{rodol}), which here we write as%

\begin{equation}
\mathcal{R}_{\mathbf{a}}-\frac{1}{2}\varepsilon^{\mathbf{a}}(R+\Lambda
)=\mathcal{T}_{\mathbf{a}} \label{eh8}%
\end{equation}
where
\begin{equation}
\mathcal{R}_{\mathbf{a}}=R_{\mathbf{a}}^{\mathbf{b}}\varepsilon_{\mathbf{b}%
}\in\sec%
{\displaystyle\bigwedge\nolimits^{1}}
T^{\ast}M,\mathbf{a,b=0,1,2,3} \label{eh9}%
\end{equation}
are the Ricci $1$-forms, the $R_{\mathbf{a}}^{\mathbf{b}}$ being the
components of the Ricci tensor in the $\{\mathbf{e}^{\mathbf{a}}%
\otimes\varepsilon_{\mathbf{b}}\}$ of $T_{1}^{1}M$.

\section{\textquotedblleft Energy-Momentum Conservation\textquotedblright\ and
$\Lambda$}

\textbf{48. }An equation equivalent to Eq.(\ref{eh8}) is the following
one\footnote{Details in how to obtain this equation may be found, e.g., in
\cite{rodol}.}:%
\begin{equation}
-d\star S^{\mathbf{a}}=\star\mathcal{T}^{\mathbf{a}}+\star t^{\mathbf{a}%
}+\Lambda\star\varepsilon^{\mathbf{a}}, \label{eh10}%
\end{equation}
where $S^{\mathbf{a}}\in\sec%
{\displaystyle\bigwedge\nolimits^{2}}
T^{\ast}M$ are the \textit{superpotentials} and the $t^{\mathbf{a}}\in\sec%
{\displaystyle\bigwedge\nolimits^{1}}
T^{\ast}M$ are the $1$-forms whose components are the energy momentum
pseudo-tensor of the gravitational field in a given gauge. If you are
interested in the explicit forms of these objects, please consult
\cite{rodol}. Here, the importance of Eq.(\ref{eh10}) is that applying the
differential operator $d$ to both sides of Eq.(\ref{eh10}) and moreover, if we
suppose \ that $\Lambda\in\sec%
{\displaystyle\bigwedge\nolimits^{0}}
T^{\ast}M$ is a scalar function instead of a constant, we get%

\begin{equation}
d(\star\mathcal{T}^{\mathbf{a}}+\star t^{\mathbf{a}}+\Lambda\star
\varepsilon^{\mathbf{a}})=-d\Lambda\wedge\star\varepsilon^{\mathbf{a}}
\label{eh11}%
\end{equation}

Eq.(\ref{eh11}) shows the existence of an \textquotedblleft energy-momentum
conservation law\textquotedblright\footnote{For the reason of the " ", please
consult \cite{rodol}.} only if $d\Lambda=0$. So, our conclusion is in
contradiction with the statement in \cite{sarfatti} which follows his Eq.(1.16).

Of course, in order to get an \textquotedblleft energy-momentum conservation
law\textquotedblright\ it is necessary to make $\Lambda$ a dynamic field and
write a complete Lagrangian for it. This will be not discussed here.

\textbf{49.} Introduction of a connection with torsion will not imply in the
automatic validity of Einstein field equations, and so the discourse based on
Eqs.(1.17) of \cite{sarfatti} is ad hoc.

\section{What is $\ell$}

\textbf{50. }Sarfatti \cite{sarfatti} originally defined the $\ell$ field by
his Eqs.(1.1) and (1.2) (see Eqs. (\ref{17bis}), (\ref{18} ) and (\ref{19})
above). As observed in \textbf{21 }Eqs.(1.1) and (1.2) imply that $\ell\in\sec
T_{1}^{1}M$, i.e., it is a vector valued form field. However, later Sarfatti
wrote: \textquotedblleft Define the $1$-form invariant curved spacetime tetrad
field as
\begin{equation}%
\begin{tabular}
[c]{|c|}\hline
$\ell=\sqrt{\frac{\hslash G}{c^{3}}}((d\theta)\phi-\theta d\phi)=\ell_{\mu
}dx^{\mu}.$\textquotedblright\\\hline
\end{tabular}
\ \ \ \tag{1.28}%
\end{equation}

\textbf{51. }Since $\theta,\phi\in\sec%
{\displaystyle\bigwedge\nolimits^{0}}
T^{\ast}M$ $\ $are functions according to Eq.(1.23) of \cite{sarfatti} we have
that the object defined by Eq.(1.28) must be a $1$-form field, i.e., $\ell
\in\sec%
{\displaystyle\bigwedge\nolimits^{1}}
T^{\ast}M$. So, the object defined by Eq.(1.28) cannot be the same $\ell$ as
the object defined in Eq.(1.1) in \cite{sarfatti}, which as we already said is
a section of to $T_{1}^{1}M$.

\textbf{52. }Besides this last observation it is necessary now to recall
\textbf{1 }and to emphasize here that the $\ell$ in Eq.(1.28) is only
\textit{one} $1$-form field. So, it cannot represent a tetrad which is a set
of \textit{four} $1$-form fields.

\textbf{53. }The conceptual error just mentioned, shows clearly that Sarfatti
did not grasped well the true mathematical meaning of the objects he uses.
Another example of our unfortunately not very polite statement is the last
formula in Eq.(1.31) of \cite{sarfatti}, namely%
\[%
\begin{tabular}
[c]{|c|}\hline
$\ell_{\mu}^{\mathbf{a}}=\ell_{\mu}\frac{P^{\mathbf{a}}L_{P}^{2}}{i\hbar}%
$\\\hline
\end{tabular}
\ \ \ \ ,
\]
which is a completely nonsequitur one, once our author declared that
\textquotedblleft the $\{P^{\mathbf{a}}/i\hslash\}$ is the Lie algebra of
spacetime translation group\textquotedblright. This is so because according to
his Eq.(1.2) for fixed $\mathbf{a}$ and fixed $\mu$ each one of the $\ell
_{\mu}^{\mathbf{a}}$ are real valued mappings, i.e., $\ell_{\mu}^{\mathbf{a}%
}:\mathbb{R}^{4}\supset\chi(U)\rightarrow\mathbb{R}$.

\textbf{54. }Besides that, also take notice that from Eq.(1.28) in
\cite{sarfatti} it follows that%

\begin{equation}
d\ell=-2\sqrt{\frac{\hslash G}{c^{3}}}d\theta\wedge d\phi\label{r1}%
\end{equation}
instead of his Eq.(1.29), where it is missing the \ $-$ signal.

\section{Quantization of Area}

\textbf{55}. Of course, given observations \textbf{49-54 }the claim of an
original contribution in \cite{sarfatti} declaring to have deduced gravity as
an emergent phenomenon cannot be taken seriously. Our statement will be
reinforced after the mathematical analysis of the topological part of
\cite{sarfatti}.

\textbf{56. }The theory of topological defects in ordered media is a well
developed subject (see. e.g., \cite{naka,nash,mermin}) and to expose our
criticisms to some topological considerations in \cite{sarfatti} we need to
recall some of its rudiments.

We suppose that an ordered medium can be regards as a region $U$ of the
spacetime manifold $M$ described by a function%
\begin{equation}
\Psi:U\rightarrow O, \label{qa1}%
\end{equation}
called \textit{order parameter}. In Eq.(\ref{qa1}) $O$ is called the
\textit{parameter space} or manifold of \textit{internal states}. The
specification of $O$\textit{ }depends on the particular theory of the field
$\Psi$.

\textbf{57. }The mapping $\Psi$ is supposed to vary continuously through $U$,
except at some isolated worldlines and some appropriate hypersurfaces, where
it is singular. These regions of lower dimensionality will be called
\textit{defects in spacetime}. We suppose that $U$ can be foliated as
$U=\mathbb{R\times}X$ where $X$ is $3$-dimensional manifold (a spacelike
hypersurface). In this case, in condensed matter physics, the order parameter
is a mapping $\left.  \Psi\right\vert _{X}:X\rightarrow O$ and the defects in
the \ `\textit{space }$X$' are points, lines, surfaces where $\left.
\Psi\right\vert _{X}$ is singular.

\textbf{58}. In the \ `model' imagined in \cite{sarfatti} the parameter space
$O$ is identified with the unit radius sphere $S^{2}$. We will not discuss if
this hypothesis is reasonable or not, let us accept it here.

\textbf{59. }For the theory to work, we need to \textit{cut out} from the
manifold $U\subset M$ the points where $\Psi$ is singular. If we admit, e.g.,
a single point defect, as it is the case imagined in \cite{sarfatti} then we
must take $U=\mathbb{R\times}X$ with $X=\mathbb{R}^{3}-\{0\}\simeq
\mathbb{R\times}S^{\prime2}$, where the $S^{\prime2}$ in the previous formula
is a space isomorphic, but distinct from the parameter space\ $S^{2}$ and
where $\{0\}$ stands for the location of the point defect in $3$-dimensional
space. \ Under these conditions the effective manifold modelling spacetime
where the theory rolls is:
\begin{equation}
U=\mathbb{R}^{2}\mathbb{\times}S^{\prime2} \label{qa'}%
\end{equation}

\textbf{60}. Before continuing we must comment that from the Physical point of
view to suppose that the condensate responsible for the existence of
gravitation has only a point defect seems to us an ad hoc assumption. Indeed
author of \cite{sarfatti} did not present a single argument for it.

\textbf{61. }To continue we write
\begin{align}
\Psi &  :U\rightarrow S^{2},\nonumber\\
X  &  \ni x\mapsto y=\Psi(x)\in S^{2}, \label{qa2}%
\end{align}
and introduce the \textit{usual} spherical coordinate
functions\footnote{Please do not confound these variables with the ones used
by Sarfatti, which are defined in his Eq.(1.23).} $(\theta,\varphi)$ on the
sphere $S^{2}$ with the usual domain, say\footnote{Of course to cover all
$S^{2}$ it is necessary to introduce complementary spherical coordinate
functions $(\theta^{\prime},\varphi^{\prime})$ covering $V_{2}\subset S^{2\ }$
and such that $V_{1}\cap V_{2}\neq\varnothing$. Of course $S^{2}\subset
V_{1}\cup V_{2}$.} $V_{1}\subset S^{2}$. We have then the following coordinate
representation of $\Psi$ in $V_{1}$
\begin{equation}
\Psi(x)=(\theta(y),\varphi(y)) \label{qa4}%
\end{equation}

\textbf{62. }Now, the `area element' of the parameter space\footnote{Please,
notice that this space $S^{2}$ has \textit{nothing} to do with any surface in
the spacetime manifold $M$.} $S^{2}$, which mathematicians call the
\textit{volume element }of $S^{2}$ is given once we use the coordinate
functions $(\theta,\varphi)$ covering $V_{1}\subset S^{2}$ by the $2$-form
$\upsilon\in\sec%
{\displaystyle\bigwedge}
T^{\ast}S^{2}$
\begin{equation}
\upsilon=\sin\theta d\theta\wedge d\varphi\label{qa5}%
\end{equation}

The area of the parameter space $S^{2}$ is then calculated as\footnote{This is
so because $S^{2}-V_{1}$ is a set of zero measure.}
\begin{equation}
\mathcal{A}_{S^{2}}=%
{\displaystyle\int\nolimits_{S^{2}}}
\upsilon=%
{\displaystyle\int}
\sin\theta d\theta\wedge d\varphi=4\pi\label{qa7}%
\end{equation}

\textbf{63}. It is then automatically quantized (joke)!

\textbf{64. }The \ `area element' $2$-form $\nu$ is, of course, closed because
all $3$-forms in $S^{2}$ are null.

\textbf{65. }Now, the pullback of $\upsilon$ under the mapping $\Psi$ is the
$2$-form $%
\mbox{\boldmath{$\upsilon$}}%
\in\sec%
{\displaystyle\bigwedge\nolimits^{2}}
T^{\ast}U\subset\sec%
{\displaystyle\bigwedge\nolimits^{2}}
T^{\ast}M$ such that%
\begin{equation}%
\mbox{\boldmath{$\upsilon$}}%
=\sin(\Psi^{\ast}\theta)d(\Psi^{\ast}\theta)\wedge d(\Psi^{\ast}\varphi),
\label{qa7a}%
\end{equation}
with%
\begin{equation}
(\Psi^{\ast}\theta,\Psi^{\ast}\varphi)=(\theta\circ\Psi,\varphi\circ\Psi)
\label{qa7b}%
\end{equation}
Now, let us restrict our considerations to $\mathit{\Psi}$, which is the
restriction of the mapping $\left.  \Psi\right\vert _{X}$ to $S^{\prime2}$,
i.e., the mapping%
\[
\mathit{\Psi}=\left.  \left.  \Psi\right\vert _{X}\right\vert _{S^{\prime2}%
}:X\supset S^{\prime2}\rightarrow O
\]

The integral $%
{\displaystyle\int\nolimits_{S^{\prime2}}}
\mbox{\boldmath{$\upsilon$}}%
$ such that
\begin{equation}
\mathit{\Psi}(S^{\prime2})=S^{2} \label{qa8}%
\end{equation}
is given by (see, e.g., \cite{choquet,fel})
\begin{equation}%
{\displaystyle\int\nolimits_{S^{\prime2}}}
\mbox{\boldmath{$\upsilon$}}%
=\deg(\mathit{\Psi})%
{\displaystyle\int\nolimits_{S^{2}}}
\upsilon=4\pi\deg(\mathit{\Psi}) \label{qa9}%
\end{equation}
where $\deg(\mathit{\Psi})$ denotes the \textit{Brouwer} \textit{degree} of
mapping $\mathit{\Psi}$, which we recall is the \textit{restriction} of the
mapping $\left.  \Psi\right\vert _{X}:X\rightarrow S^{2}$ to $S^{\prime2}$.
The Brouwer degree is an integer that is roughly speaking the number of times
that each point $y\in S^{2}$ is covered by the image of $S^{\prime2}$ under
$\mathit{\Psi}$, each covering counted positively or negatively depending on
the orientation of $\mathit{\Psi}$ in an open set of the point $x=\mathit{\Psi
}^{-1}(y)$. This is one of the doors from where homotopy theory makes its
entrance in Physics.

\textbf{ 66. }The $2$-form $%
\mbox{\boldmath{$\upsilon$}}%
$ is closed, but since it is defined in $U=\mathbb{R}^{2}\mathbb{\times
}S^{\prime2}$ (and thus not diffeomorphic to $\mathbb{R}^{4})$ it is not
exact. Then it must have period integrals according to de Rham theorem (see,
e.g.,\cite{choquet,naka}). Looking at Eq.(\ref{qa9}) we see that this is
indeed necessary, for otherwise, if we could write globally $%
\mbox{\boldmath{$\upsilon$}}%
=d\mathfrak{A}$, $A\in\sec%
{\displaystyle\bigwedge^{1}}
T^{\ast}U\subset\sec%
{\displaystyle\bigwedge^{1}}
T^{\ast}M$ then Stokes theorem would give
\begin{equation}
\int_{S^{\prime2}}%
\mbox{\boldmath{$\upsilon$}}%
=\int_{S^{\prime2}}d\mathfrak{A}=\int_{\partial S^{\prime2}}\mathfrak{A}%
=\int_{\varnothing}\mathfrak{A}=0, \label{qa10}%
\end{equation}
contradicting Eq.(\ref{qa9}).

\textbf{67.} We end our observations by remarking that in \cite{sarfatti} the
$2$-form written
\begin{equation}
A=\sqrt{\frac{\hbar G}{c^{3}}}dl=2\frac{\hbar G}{c^{3}}d\theta\wedge
d\varphi\tag{1.29}%
\end{equation}
is supposed to be an area \ `flux density'. Of course, if $(\theta,\varphi)$
are interpreted as spherical coordinate functions for $S^{2}$ this is not
true, because, in order to be the $2$-form\ `area element' of a sphere a
factor $\sin\theta$ is missing. However, introduce coordinate
functions\footnote{We leave aside the dimensional factor $\sqrt{\frac{\hbar
G}{c^{3}}}$ in what follows.} $(a,b)$ on $S^{2}$ covering $V_{1}\subset S^{2}$
and such that
\begin{equation}
a=-\cos\theta,\text{ }b=\varphi\label{qa11}%
\end{equation}

Then, we have immediately from Eq.(\ref{qa5}) that
\begin{equation}
\upsilon=-2da\wedge db \label{qa12}%
\end{equation}
which in $V_{1}$ can be written as the differential of the $1$-form field
$\mathfrak{a}_{V_{1}}\in\sec%
{\displaystyle\bigwedge\nolimits^{1}}
T^{\ast}V_{1}\subset\sec%
{\displaystyle\bigwedge\nolimits^{1}}
T^{\ast}S^{2}$ such that%
\begin{equation}
\mathfrak{a}_{V_{1}}=bda-adb \label{qa13}%
\end{equation}

\textbf{68.} Now, consider again $%
\mbox{\boldmath{$\upsilon$}}%
\in\sec%
{\displaystyle\bigwedge\nolimits^{2}}
T^{\ast}U\subset\sec%
{\displaystyle\bigwedge\nolimits^{2}}
T^{\ast}M$ which is the pullback under the mapping $\mathit{\Psi}$ \ of
$\upsilon\in\sec%
{\displaystyle\bigwedge\nolimits^{2}}
T^{\ast}S^{2}$.

\textbf{69. }Introduce Cartesian and spherical coordinate functions in
$X=\mathbb{R}^{3}-\{0\}=\mathbb{R\times S}^{\prime2}$ with center in $\{0\}$,
the defect localization point. The spherical coordinates ($r,%
\mbox{\boldmath{$\theta$}}%
,%
\mbox{\boldmath{$\varphi$}}%
$) on $S^{\prime2}$ are given by the restrictions of the functions $(%
\mbox{\boldmath{$\theta$}}%
,%
\mbox{\boldmath{$\varphi$}}%
)$ on $S^{%
\acute{}%
2}$. We now specify the restriction of the \textit{coordinate representation}
of the mapping $\mathit{\Psi}$ as functions of the spherical coordinates $(%
\mbox{\boldmath{$\theta$}}%
,%
\mbox{\boldmath{$\varphi$}}%
)$, i.e., we write:%
\begin{equation}
\mathit{\Psi}(%
\mbox{\boldmath{$\theta$}}%
,%
\mbox{\boldmath{$\varphi$}}%
)=\left[
\begin{array}
[c]{c}%
\mathit{\Psi}_{1}(%
\mbox{\boldmath{$\theta$}}%
,%
\mbox{\boldmath{$\varphi$}}%
)\\
\mathit{\Psi}_{2}(%
\mbox{\boldmath{$\theta$}}%
,%
\mbox{\boldmath{$\varphi$}}%
)\\
\mathit{\Psi}_{3}(%
\mbox{\boldmath{$\theta$}}%
,%
\mbox{\boldmath{$\varphi$}}%
)
\end{array}
\right]  =\left[
\begin{array}
[c]{c}%
\sin%
\mbox{\boldmath{$\theta$}}%
\cos n%
\mbox{\boldmath{$\varphi$}}%
\\
\sin%
\mbox{\boldmath{$\theta$}}%
\sin n%
\mbox{\boldmath{$\varphi$}}%
\\
\cos%
\mbox{\boldmath{$\theta$}}%
\end{array}
\right]  , \label{qa13a}%
\end{equation}
with $n\in\mathbb{Z}$. This means that%
\begin{equation}
(\theta,\varphi)=(%
\mbox{\boldmath{$\theta$}}%
,n%
\mbox{\boldmath{$\varphi$}}%
)=(\mathit{\Psi}^{\ast}\theta,\mathit{\Psi}^{\ast}\varphi). \label{qa14}%
\end{equation}

\bigskip We remark that as defined $\mathit{\Psi}$ is a smooth mapping outside
the poles. Moreover, $\mathit{\Psi}$ maps $S^{^{\prime}2}$ n times around
$S^{2}$. This is easily seem if we observe from Eq.(\ref{qa13a})
that\ $\mathit{\Psi}$ maps any circle $%
\mbox{\boldmath{$\theta$}}%
=%
\mbox{\boldmath{$\theta$}}%
_{0}$ $n$ times on the corresponding circle in $S^{2}.$

\ \ \ \ \ \textbf{70.} Write moreover as usual the following relations between
the Cartesian and spherical coordinate functions on $X=\mathbb{R}^{3}-\{0\}$,%
\begin{align}
\mathtt{x}  &  =r\sin%
\mbox{\boldmath{$\theta$}}%
\cos%
\mbox{\boldmath{$\varphi$}}%
\text{, \texttt{y}}=r\sin%
\mbox{\boldmath{$\theta$}}%
\cos%
\mbox{\boldmath{$\varphi$}}%
\text{, }\mathtt{z}=r\cos%
\mbox{\boldmath{$\theta$}}%
,\nonumber\\
r  &  =\sqrt{\mathtt{x}^{2}+\mathtt{y}^{2}+\mathtt{z}^{2}}. \label{qa15}%
\end{align}

Then the non exact $2$-form $%
\mbox{\boldmath{$\upsilon$}}%
\in\sec%
{\displaystyle\bigwedge\nolimits^{2}}
T^{\ast}U\subset\sec%
{\displaystyle\bigwedge\nolimits^{2}}
T^{\ast}M$ can be described as the differential of the following two $1$-form
fields on the regions $U_{1}$ and $U_{2}$,%
\begin{align}
\mathfrak{A}_{1}  &  =-n(\cos%
\mbox{\boldmath{$\theta$}}%
-1)d%
\mbox{\boldmath{$\varphi$}}%
\text{ on }U_{1}=\mathbb{R\times(R}^{3}-\{r=0\text{ or \texttt{z}%
}<0\}),\nonumber\\
\mathfrak{A}_{2}  &  =-n(\cos%
\mbox{\boldmath{$\theta$}}%
+1)d%
\mbox{\boldmath{$\varphi$}}%
\text{ on }U_{2}=\mathbb{R\times(R}^{3}-\{r=0\text{ or \texttt{z}%
}>0\}),\label{qa16}\\
U_{1}\cap U_{2}  &  =S^{1}\text{ .}%
\end{align}

If we write the representatives of the $2$-form $%
\mbox{\boldmath{$\upsilon$}}%
$ on the same regions $U_{1}$ and $U_{2}$ as $%
\mbox{\boldmath{$\upsilon$}}%
_{1}$ and $%
\mbox{\boldmath{$\upsilon$}}%
_{2}$ we can write
\begin{align}
\int_{S^{\prime2}}%
\mbox{\boldmath{$\upsilon$}}%
&  =\int_{U_{1}}%
\mbox{\boldmath{$\upsilon$}}%
_{1}+\int_{U_{2}}%
\mbox{\boldmath{$\upsilon$}}%
_{2}\nonumber\\
&  =\int_{S^{1}}(\mathfrak{A}_{1}-\mathfrak{A}_{2})=n\int_{S^{1}}d%
\mbox{\boldmath{$\varphi$}}%
=4\pi n \label{QA18}%
\end{align}

\textbf{71. }Finally we can write on $U_{1}\cup U_{2}-\{$worldline of the
defect$\}$
\begin{align}%
\mbox{\boldmath{$\upsilon$}}%
&  =n\sin%
\mbox{\boldmath{$\theta$}}%
d%
\mbox{\boldmath{$\theta$}}%
\wedge d%
\mbox{\boldmath{$\varphi$}}%
\nonumber\\
&  =\frac{n}{r}\left(  \mathtt{x}d\mathtt{y}\wedge d\mathtt{z}+\mathtt{y}%
d\mathtt{z}\wedge d\mathtt{x}+\mathtt{z}d\mathtt{x}\wedge d\mathtt{y}\right)
, \label{qa19}%
\end{align}

\textbf{72. }Readers that know the $U(1)$ principal fiber bundle formulation
of the \textit{magnetic monopole} will recognize that apart for the correct
physical units $%
\mbox{\boldmath{$\upsilon$}}%
$ describes the field of a magnetic monopole. This is not a coincidence, of
course, since the formulation of both problems (the point defect and the
monopole one) have many common ingredients \footnote{See, e.g., \cite{choquet}%
.}.

\section{Conclusion}

Sarfatti's paper\footnote{More specifically (as said in the abstract) the
first version posted at the arXiv.
\par
{}} \cite{sarfatti}, we regret to say, is unfortunately a potpourri of
nonsense Mathematics\footnote{This paper has been written as a consequence of
an exercise that we proposed to some of our students. Eventually, Sarfatti
will not like it, and will probably say that we are very pedantic, but
eventually (we hope) he will use it to write a better version of his paper. In
any case, we would like that he be aware that in writing it we found also
inspiration in Aristotle [who in his Nicomachean Ethics, book 1, Chapter 6
said in a similar situation where he could not agree with the presentations of
some of his friends on a \ given subject that: \ `...piety requires us to
honor truth above our friends'], and also in our (late) friend Pertti Lounesto
that enlightened us for many years with his posters on errors and
counterexamples to \ `theorems' found in the literature on Clifford
algebras.}. The fact that he found endorsers which permitted him to\ put his
article in the arXiv is a preoccupying fact. Indeed, the incident shows that
endorsers did not pay attention to what they read, or worse, that there are a
lot of people with almost null mathematical knowledge publishing
Physics\footnote{And also mathematical papers, as e.g., \cite{car}. See oour
analysis of that bad paper published in Nonlinear Analysis in
\cite{rodoliv2006}.
\par
.} papers replete of nonsense Mathematics. We recall here that among others,
author of \cite{sarfatti} confounded a single $1$-form field (the one given by
Eq.(1.28)) with (non trivial part) a tetrad, which is a set of \textit{four}
distinct $1$-form fields, wrote in a wrong and misleading way the
Einstein-Hilbert Lagrangian density, misleads the real nature of the
connection $1$-forms, wrote a misleading \ `conservation' equation to deduce
that the cosmological constant need not be a constant in General Relativity,
supposed in an ad hoc way that Einstein's equations also holds in a spacetime
with torsion, and finally, used in a misleading way topological arguments.
Also that author did not leave it clear what are the hypotheses he used. A
careful reading of \cite{sarfatti} shows that his hypotheses are completely ad
hoc assumptions, since in our view no arguments from Physics or Mathematics
are given for them. Summing up, we must say that Sarfatti's claim to have
deduced Einstein's equations as an emergent phenomena is an statement that
cannot be taken seriously.


\begin{thebibliography}{99}                                                                                               %


\bibitem {car}Carvalho, L. A. V., A Comment \textquotedblleft On Some
Contradictory Computations in Multi-Dimesional Mathematics\textquotedblright,
\textit{Nonlinear Analysis} \textbf{63}, 725-734 (2005).

\bibitem {collins}Borowski, E. J. and Borwein, J. M., \textit{The Collins
Dictionary of Mathematics}, Harper Collins Publ., Glasgow, 1989.

\bibitem {choquet}Choquet-Bruhat, Y., DeWitt-Morette, C., and Dillard-Bleick,
M., \textit{Analysis, Manifolds and Physics} (revised edition), North-Holland
Publ. Co., Amsterdam, 1977.

\bibitem {fel}Felsager, B., \textit{Geometry, Particles and Fields}, Springer,
Berlin, 1998.

\bibitem {kono}Kobayashi, \ S. and Nomizu, K., \textit{Foundations of
Differential Geometry}, vol. I, Interscience Publ., New York, 1963.

\bibitem {naka}Nakahara, M., \textit{Geometry, Topology and Physics},
Institute of Physics Publ., Bristol,1990.

\bibitem {nash}Nash, C. and Sen, S., \textit{Topology and Geometry for
Physicists}, Academic Press, New York, 1983.

\bibitem {mermin}Mermin, N. D., The Topological Theory of Defects in Ordered
Media, \textit{Rev. Mod. Phys. }\textbf{51}\textit{,} 591-648 (1979).

\bibitem {moro}Mosna, R. A. and Rodrigues, W .A. Jr., The Bundles of Algebraic
and Dirac-Hestenes Spinor Fields, \textit{J. Math. Phys}.\ \textbf{45},
2945-2966 (2004). [math-phys/]

\bibitem {olro}Rodrigues, W. A. Jr. and Oliveira, E. Capelas, Clifford Valued
Differential Forms, and Some Issues in Gravitation, Electromagnetism and
\textquotedblleft Unified Theories, \textit{Int. J. Mod. Phys. D. }%
\textbf{13}, 1897-1915 (2004).

\bibitem {rodol}Rodrigues, W. A. Jr. and Oliveira, E. Capelas, \textit{The
Many Faces of Maxwell, Dirac and Einstein Equations. A Clifford Bundle
Approach}, RP 56/05 IMECC-UNICAMP
[{\small http://www.ime.unicamp.br/rel\_pesq/2005/rp56-05.html].}

\bibitem {roqui}Rodrigues, W. A. Jr. and Souza, Q. A. G., An Ambigous
Statement Called \ `Tetrad Postulate' and the Correct Field Equations
Satisfied by the Tetrad Fields, \textit{Int. J. Mod. Phys. D}. \textbf{14},
2095-2150 (2005). {\small [math-ph/0411085]}

\bibitem {rodoliv2006}Olivieira, E. Capelas, and Rodrigues, W. A., Jr.,
\textit{A Comment \textquotedblleft On Some Contradictory Computations in
Multi-Dimesional Mathematics}\textquotedblright. math.GM/0603599

\bibitem {rovelli}Rovelli, C., \textit{Quantum Gravity}, Cambridge University
Press, Cambridge, 2004.

\bibitem {sarfatti}Sarfatti, J., Emergent Gravity {\small [gr-qc/0602022]}
\end{thebibliography}
\end{document}